\documentclass{IEEEoj}
\usepackage{cite}
\usepackage{amsmath,amssymb,amsfonts}
\usepackage{algorithmic}
\usepackage{graphicx,color}
\usepackage{caption}
\usepackage{subcaption}
\usepackage{textcomp}
\usepackage{multicol}
\usepackage{stfloats}
\usepackage[T1]{fontenc}
\usepackage{float}
\usepackage{orcidlink}
\usepackage{newtxmath} 
\def\BibTeX{{\rm B\kern-.05em{\sc i\kern-.025em b}\kern-.08em
    T\kern-.1667em\lower.7ex\hbox{E}\kern-.125emX}}
\AtBeginDocument{\definecolor{ojcolor}{cmyk}{0.93,0.59,0.15,0.02}}
\def\OJlogo{\vspace{-14pt}\includegraphics[height=28pt]{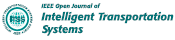}}
\begin{document}
\receiveddate{20 February, 2025}
\reviseddate{XX Month, XXXX}
\accepteddate{XX Month, XXXX}
\publisheddate{XX Month, XXXX}
\currentdate{XX Month, XXXX}
\doiinfo{OJITS.2022.1234567}

\title{Identification and Classification of Human Performance related Challenges during Remote Driving}

\author{OLE HANS\textsuperscript{1,2}\orcidlink{0009-0009-0239-9469} and JÜRGEN ADAMY\textsuperscript{1}\orcidlink{0000-0001-5612-4932}}
\affil{Department of Control Methods and Intelligent Systems, Technical University of Darmstadt, Darmstadt, 64289 Germany}
\affil{Department of Operational Safety, Vay Technology GmbH, Berlin, 
12099 Germany}
\corresp{CORRESPONDING AUTHOR: OLE HANS (e-mail: ole.hans@vay.io).}
\authornote{The work of Ole Hans was supported by the Operational Safety Department of Vay Technology GmbH.}
\markboth{Identification and Classification of Human Performance related Challenges during Remote Driving}{Ole Hans and Jürgen Adamy}

\begin{abstract}
Remote driving of vehicles, where human operators directly control vehicles from distance, is gaining in importance in the transportation sector, especially when Automated Driving Systems (ADSs) reach the limits of their Operational Design Domain (ODD). This study investigates the challenges faced by human Remote Drivers (RDs) during remote driving operations, particularly focusing on the identification and classification of human performance-related challenges through a comprehensive analysis of remote driving data from real-world urban ODDs in Las Vegas. For this purpose, a total of 183 RD performance-related Safety Driver (SD) interventions were analyzed and classified using an introduced severity classification developed specifically for remote driving applications. As it is essential to prevent the need for SD interventions for a safe market launch of the remote driving technology, this study identified and analyzed harsh driving events to detect an increased likelihood of interventions by the SD. In addition, the results of the subjective RD questionnaire are used to evaluate whether the objective metrics from SD interventions and harsh driving events can also be confirmed by the RDs and whether additional challenges can be uncovered. The analysis reveals learning curves, showing a significant decrease in SD interventions as RD experience increases. Early phases of remote driving experience, especially below 200 km of experience, showed the highest frequency of safety-related events, including braking late for traffic signs and responding impatiently to other traffic participants. Over time, RDs follow defined rules for improving their control, with experience leading to less harsh braking, acceleration, and steering maneuvers. The study contributes to understanding the requirements of RDS, emphasizing the importance of targeted training to address human performance limitations and optimize remote driving safety. It further highlights the need for system improvements to address challenges like latency and the limited haptic feedback replaced by visual feedback, which affect the RDs' perception and vehicle control.
\end{abstract}

\maketitle
\section{INTRODUCTION}
\label{sec_introduction}
\IEEEPARstart{R}{emote} operation is becoming increasingly important in the transportation sector. In particular, remote driving is being developed as a supporting technology for Automated Driving Systems (ADSs) when these reach the limits of their Operational Design Domain (ODD). By combining the technological capabilities of ADS with the flexibility and intuition of human operators, remote driving can thus serve as a bridge toward a safer incorporation of ADS in urban environment \cite{hans2024backedautonomy}. While technological advances are improving the control of remote-controlled vehicles, human control of Remote Driving Systems (RDSs) remains a relevant factor for the safety, controllability and efficiency of such RDSs. 
With the increasing integration of remote driving technologies in increasingly diverse environments such as urban, industrial, and highway scenarios, the need for research into the practical implementation and the associated challenges for Remote Drivers (RDs) is also growing. Institutions such as the British Standards Institution (BSI), which has published guidelines for remote driving and its human \mbox{factors \cite{BSIFlex1887}}, and the german Federal Highway Research Institute (BASt), which has identified the need for research in the field of teleoperation \cite{TeleoperationReport}, underline the need for further scientific studies. In addition, reports from management consultancies such as McKinsey \cite{Kelkar2025, Heineke2024} highlight the growing need and increasing demand for remote driving in consumer vehicles. Current regulatory efforts to introduce a legal basis for the operation of remotely driven vehicle in Germany \cite{roedl_fernlenkverordnung_2024} also emphasize the relevance of further research in this area. In particular, the focus is on issues related to vehicle control and human-machine interaction and operational safety. Despite this need, much research work to date has been based on simulations \cite{neumeier2019teleoperation} or was executed in controlled \mbox{environments \cite{den2022design, tener2022driving}}, which provide valuable insights into technical and behavior-based aspects, but do not sufficiently capture the practical challenges of real-world driving situations.

This study addresses this research gap by analyzing real-world remote driving data obtained from RDS deployed in real-world driving environments operated by a diverse set of RDs. The aim is to systematically record and evaluate the challenges for human RDs. This is done based on a combination of objective field data and subjective assessments of professionally trained RDs from a remote driving car sharing service. Therefore, \mbox{Safety Driver (SD)} interventions, harsh driving events as potential precursors of these interventions as well as the subjective feedback of the RDs are analyzed. This multidimensional approach identifies key challenges, but also reveals possible discrepancies between theoretical assumptions and practical implementation. In addition, the study makes a contribution to the validation or refutation of existing simulation results by using real-world remote driving data as an empirical basis. This can provide valuable insights for the design of future RDSs, particularly with regard to training strategies, Human-Machine Interfaces (HMIs) and the development of safety mechanisms. Through the practical analysis, this study offers new scientifically insights into the challenges and optimization possibilities of remote driving and thus contributes to the further development of safe and efficient remote control technologies.

In Section \ref{sec_relatedwork}, this study provides literature research relevant to the study, and \mbox{Section \ref{sec_methodology}} defines the methodological approach used in this study. \mbox{Section \ref{sec_systemmodel}} describes the RDS in its components and the respective ODD as the basis for this study. The results of the in total 183 analyzed SD interventions are described in \mbox{Section \ref{sec_SD_interventions_results}} using a proposed SD intervention severity definition to classify the severity of these disengagements. As it is essential for a safe market launch of remote driving technology to prevent the need for SD interventions, this study identifies potentially safety-relevant driving situations at an early stage. In particular, harsh driving events are identified and analyzed to detect an increased likelihood of disengagement by the RD in \mbox{Section \ref{sec_harsh_driving_events}}. In addition, the results of the subjective RD questionnaire are described in \mbox{Section \ref{sec_results_questionaire}} to evaluate whether the objective metrics from SD interventions and harsh driving events can also be confirmed by the RDs and whether additional challenges can be uncovered. The limitations of the results are outlined in \mbox{Section \ref{sec_limitation}}. Finally, the conclusion and potential further work are presented in \mbox{Section \ref{sec_conclusion}}. 

\section{RELATED WORK}
\label{sec_relatedwork}
This Section presents an overview of research related to the remote driving technology, their limitations and the measurement of driving performance.

\subsection{Remote Driving}
\label{sec_remotedriving}
\begin{figure}[ht]
\centering
  \includegraphics[width=0.3\textwidth]{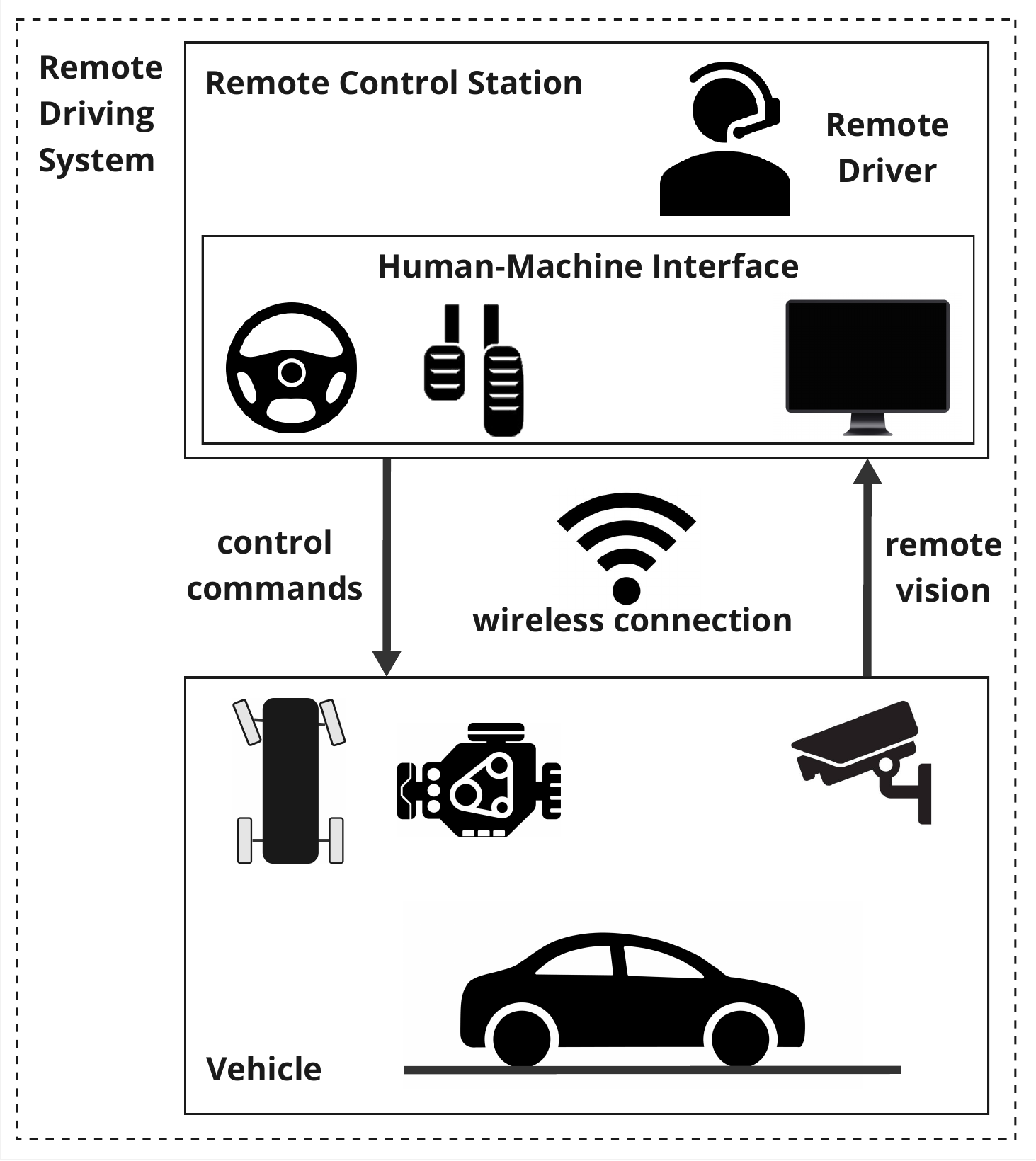}
    \caption{Simplified visualization of a Remote Driving System (RDS).}
    \label{Fig_RDSOverview}
    \vspace{-3mm}
\end{figure}

Remote driving is a part of the remote operation concepts, which also include concepts like remote assistance and remote \mbox{monitoring \cite{UNECE}}. Based on fundamental concept work by \mbox{Bogdoll et al. \cite{bogdoll2022taxonomy}}, which corresponds to established classifications for ADSs \cite{SAEI}, it can be further distinguished between direct and indirect control. During direct control the RD continuously monitors and controls the vehicle anytime. In case of indirect control concepts the RD only intervenes when necessary e.g. if the ADS reaches its ODD \mbox{boundaries \cite{majstorovic2022survey}}. 
The remote driving technology relies on a solid wireless connection between the Remote Control \mbox{Station (RCS)}, as shown in \mbox{Fig. \ref{Fig_RDSOverview}}, and the vehicle, which allows for real-time monitoring and control. This connection enables the RD to remotely access portions of the ODD using data transmitted via cameras and sensors, but with a delay that degrades real-time situational awareness of the \mbox{RD \cite{kettwich2021teleoperation, neumeier2019teleoperation}}. 

RDSs face both technical and human challenges, where technical limitations such as latency or reduced video quality affect the RD performance. At the same time, the limited haptic feedback leads to reduced situational awareness as the RD relies solely on visual feedback or synthetic haptic feedback, which can limit the RD's ability to control the vehicle \cite{chen2007human, hortal2019rehabilitation, hans2024human}. Perceptual distortions, e.g. due to scale ambiguity or unnatural camera perspectives, also impair spatial orientation and make decision-making more \mbox{difficult \cite{chen2007human, fong2001vehicle}}. Latency and video quality constraints the RD to consistently perceive the driving situation and impair the responsiveness of the RD \cite{tener2022driving, neumeier2019teleoperation}. Stable transmission of sensor data via mobile networks poses a further challenge, as fluctuations in data quality can affect the controllability of the vehicle \cite{georg2020sensor}. In addition, an intuitive representation of the vehicle environment is required to avoid misinterpretations and delays in decision-making \cite{georg2019adaptable}.  

In addition, Hans et al. \cite{hans2023operational} strongly emphasize that the definition of the ODD must not only take into account technical system limits, but also human performance in order to ensure the functionality of RDS. Overcoming these challenges requires both technological developments, especially to reduce latency and improve sensor integration, as well as targeted training measures to optimize human performance in remote driving \cite{Schwindt-Drews2024, hans2024human}.

\subsection{Driving Performance Measuring}
\label{sec_drivingperformancemetrics}
The driving style of a driver describes the operation of the vehicle controls in the context of the driving scene external conditions such as time of day or weather \cite{martinez2017driving}. It can be classified on the basis of behavioral and vehicle dynamic characteristics, including speed selection, acceleration patterns and risk-taking behavior \cite{elander1993behavioral}. Three driving styles are often distinguished into comfort-oriented, moderate and dynamic/aggressive driving \cite{bellem2018comfort, abendroth2009leistungsfahigkeit}. Several authors used these categories and note that other definitions can fit within this \mbox{framework \cite{buyukyildiz2017identification, karjanto2017simulating, martinez2017driving, schulz2008analyse}}. Other studies confirm the correlation between driving style and vehicle dynamics, with aggressive drivers exhibiting higher longitudinal and lateral \mbox{acceleration \cite{pion2012fingerprint}}. Driving style recognition is based on a variety of metrics, although there is no uniform standard in literature, partly due to the diverse applications of driving style measurements. Depending on the application, e.g. driver correction, fuel consumption optimization or safety enhancement, different parameters such as speed, acceleration, deceleration and braking patterns are used \cite{corti2013quantitative, manzoni2010driving, guardiola2014modelling}. 

\begin{table}[ht]
\centering
\begin{tabular}{lll}
\hline
\textbf{Event Type} & \textbf{Threshold} & \textbf{Reference} \\ \hline \hline
Braking events       & $< -2.5 \, \text{m/s}^2$       & \cite{schulz2008analyse, buyukyildiz2017identification, schwab2019methode, lee2024study} \\ \hline
Acceleration events  & $> 2.0 \, \text{m/s}^2$       & \cite{karjanto2017simulating, mayser2004fahrerassistenzsysteme, lee2024study, svensson2015tuning, bosetti2014human} \\ \hline
Right steering events & $> 2.9 \, \text{m/s}^2$      & \cite{hugemann2003longitudinal, dorr2014online, buyukyildiz2017identification, schwab2019methode} \\ \hline
Left steering events  & $< -2.9 \, \text{m/s}^2$     & \cite{hugemann2003longitudinal, dorr2014online, buyukyildiz2017identification, schwab2019methode} \\ \hline
Deceleration Jerk events & $< -0.9 \, \text{m/s}^3$ & \cite{murphey2009driver, svensson2015tuning} \\ \hline
Acceleration Jerk events & $> 0.9 \, \text{m/s}^3$  & \cite{kilinc2012determination, bae2019toward} \\ \hline
Right lateral jerk events & $> 0.9 \, \text{m/s}^3$ & \cite{bae2019toward} \\ \hline
Left lateral jerk events  & $< -0.9 \, \text{m/s}^3$ & \cite{bae2019toward} \\ \hline
\end{tabular}
\caption{Vehicle motion parameters, thresholds and corresponding references for identifying harsh driving events.}
\label{motionparajournal}
\end{table}

Martinez et al. \cite{martinez2017driving} provide a comprehensive overview of relevant metrics derived from real-world driving data. The data used to choose these driving metrics is derived from a variety of sources that investigate typical human driving patterns in real-world settings. In particular, lateral acceleration provides information on cornering behavior and vehicle \mbox{stability \cite{schulz2008analyse}}, while longitudinal dynamics provide indicators for speed and deceleration \mbox{control \cite{karjanto2017simulating}}. Advanced metrics such as jerk as the rate of change of acceleration, throttle pedal pressure and braking frequency also enable more precise detection of sudden or aggressive driving \mbox{maneuvers \cite{murphey2009driver, miyajima2007driver}}. The combination of longitudinal and lateral dynamics in addition allows a detailed and more comprehensive analysis of the driving \mbox{style \cite{doshi2010examining}}. In remote driving scenarios, previous research shows that remote driving increases the tendency to motion sickness, which requires targeted analysis of lateral and longitudinal driving \mbox{precision \cite{papaioannou2023motion}}.

\section{METHODOLOGY}
\label{sec_methodology}
To identify and classify scenarios that frequently lead to challenges in remote driving a three-stage assessment approach was used in this work. This combines objective metrics and subjective criteria for the comprehensive evaluation of safety relevant driving situations. The approach is based on the analysis of SD interventions in \mbox{Section \ref{sec_SD_interventions_results}} in which the SD overrides the control of the \mbox{RD ($n = 25$)}. As these interventions are already aimed at avoiding potentially dangerous situations, an additional analysis of harsh driving events from the \mbox{RD ($n = 20$)} was carried out in \mbox{Section \ref{sec_harsh_driving_events}}. These events can serve as indicators for situations in which an SD intervention might be necessary. The objective analysis is supplemented by a subjective evaluation based on direct feedback from \mbox{RD ($n = 19$)} in \mbox{Section \ref{sec_results_questionaire}}. This enables the consideration of driver-related perceptions and challenges in relevant driving scenarios, in addition to technically measurable factors. Based on this methodology, driving metrics form an essential basis for evaluating the performance, controllability and efficiency of RDs within an RDS. To ensure a consistent and meaningful data set, remote driving sessions with a duration of less than 0.2 minutes were excluded to avoid incomplete or irrelevant data points due to vehicle start tests performed. In addition, data recorded under specific test conditions of the RDS were removed to prevent distortion of the data. Furthermore, only remote driving within the defined ODD from \mbox{Section IV.\ref{sec_RDS_ODD}} were taken into account. 

All RDs completed the standardized remote driving training program from \mbox{Section IV.\ref{sec_RDTraining}} in advance, with the exception of those who are still in the training phase. This ensures a uniform level of competence so that the influence of cumulative remote driving experience on driving performance could be investigated without differences in basic driving skills acting as a confounding factor. All RDs included had a valid US driving license and at least two years of conventional driving experience.

\begin{figure*}
    \centering
    \begin{subfigure}{0.32\textwidth}
        \includegraphics[width=\linewidth]{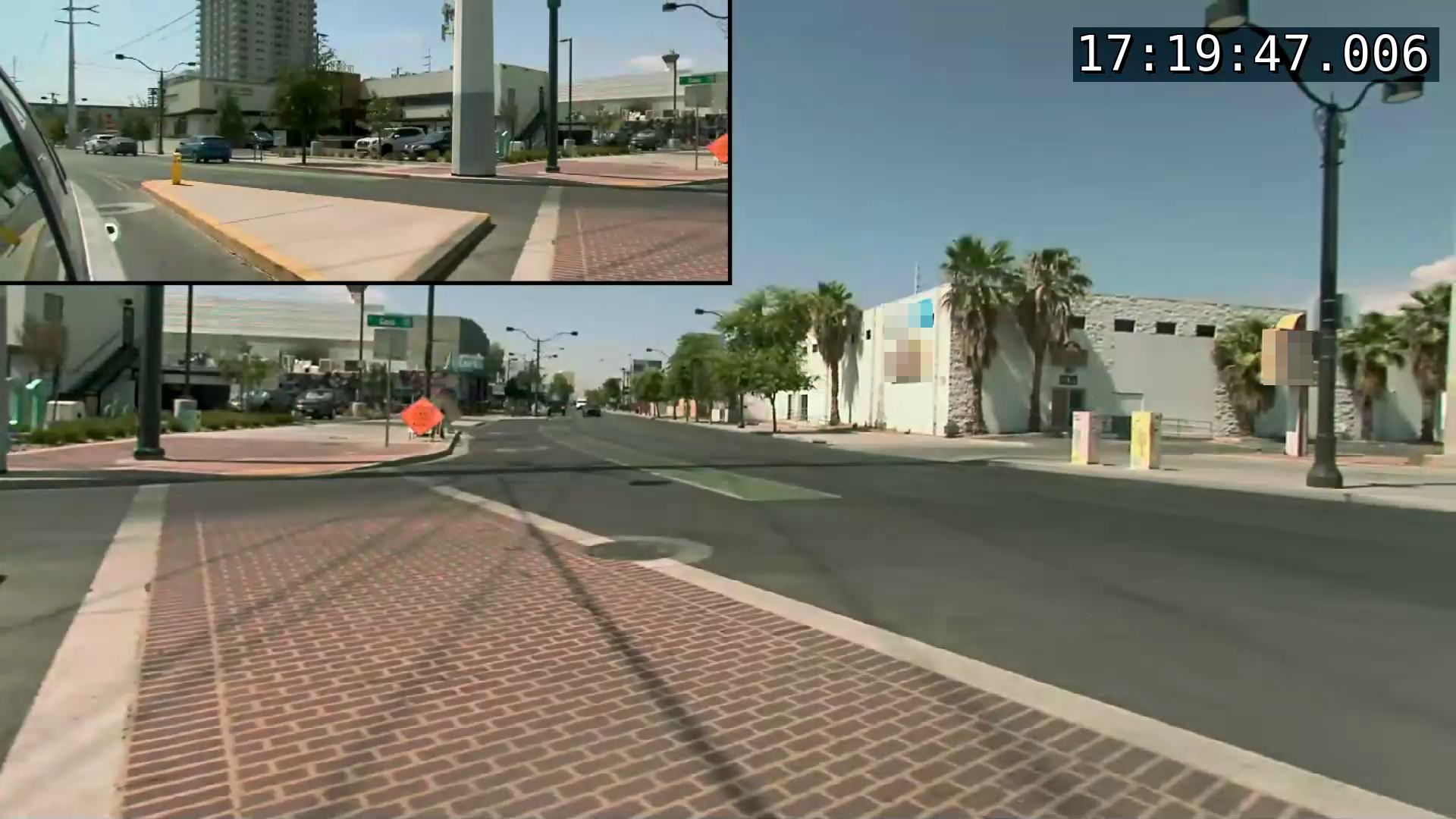}
        \caption{Left side screen}
        \label{Left_side_screen}
    \end{subfigure}
    \hfill
    \begin{subfigure}{0.32\textwidth}
        \includegraphics[width=\linewidth]{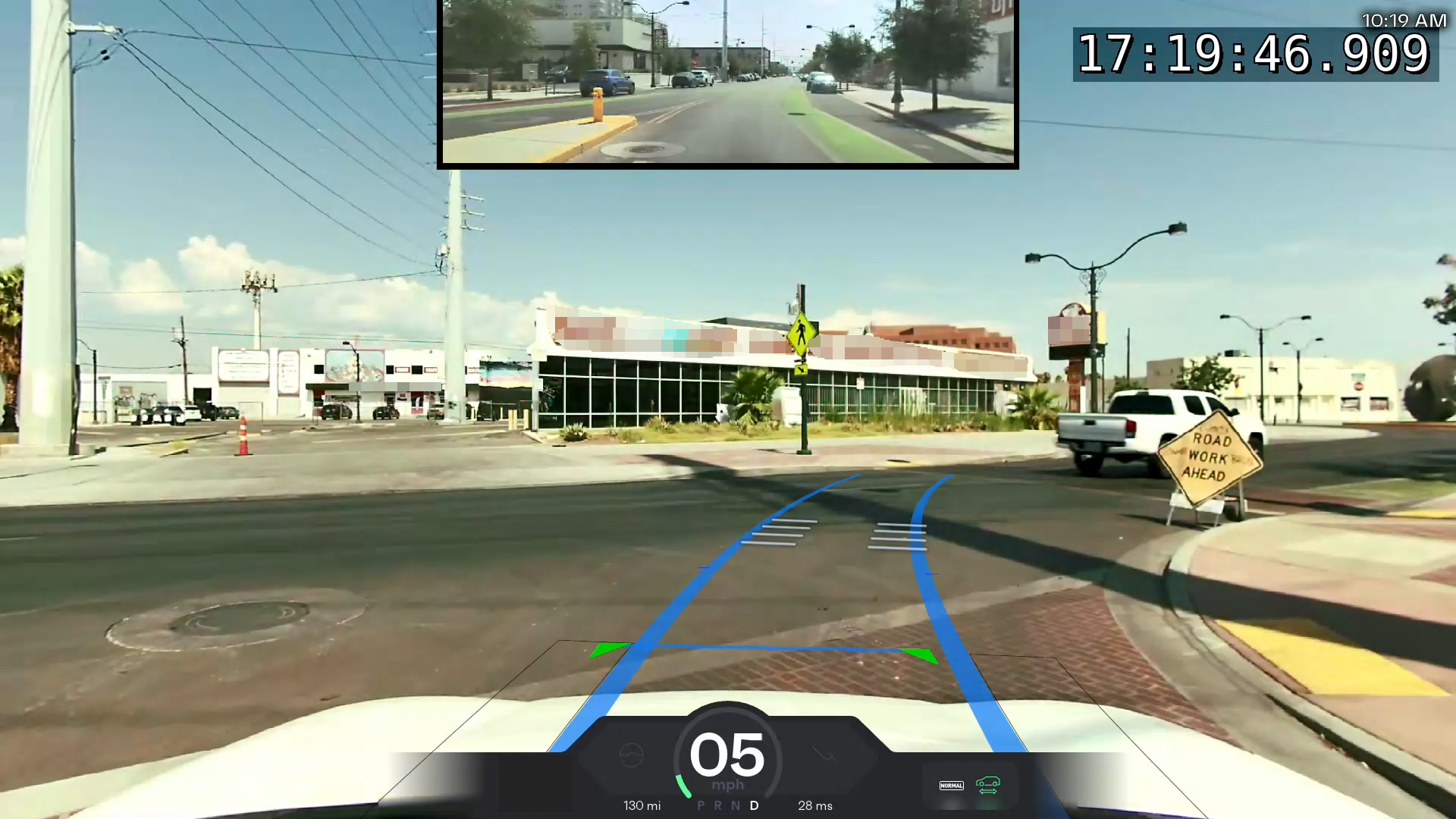}
        \caption{Front screen}
        \label{Front_screen}
    \end{subfigure}
    \hfill
    \begin{subfigure}{0.32\textwidth}
        \includegraphics[width=\linewidth]{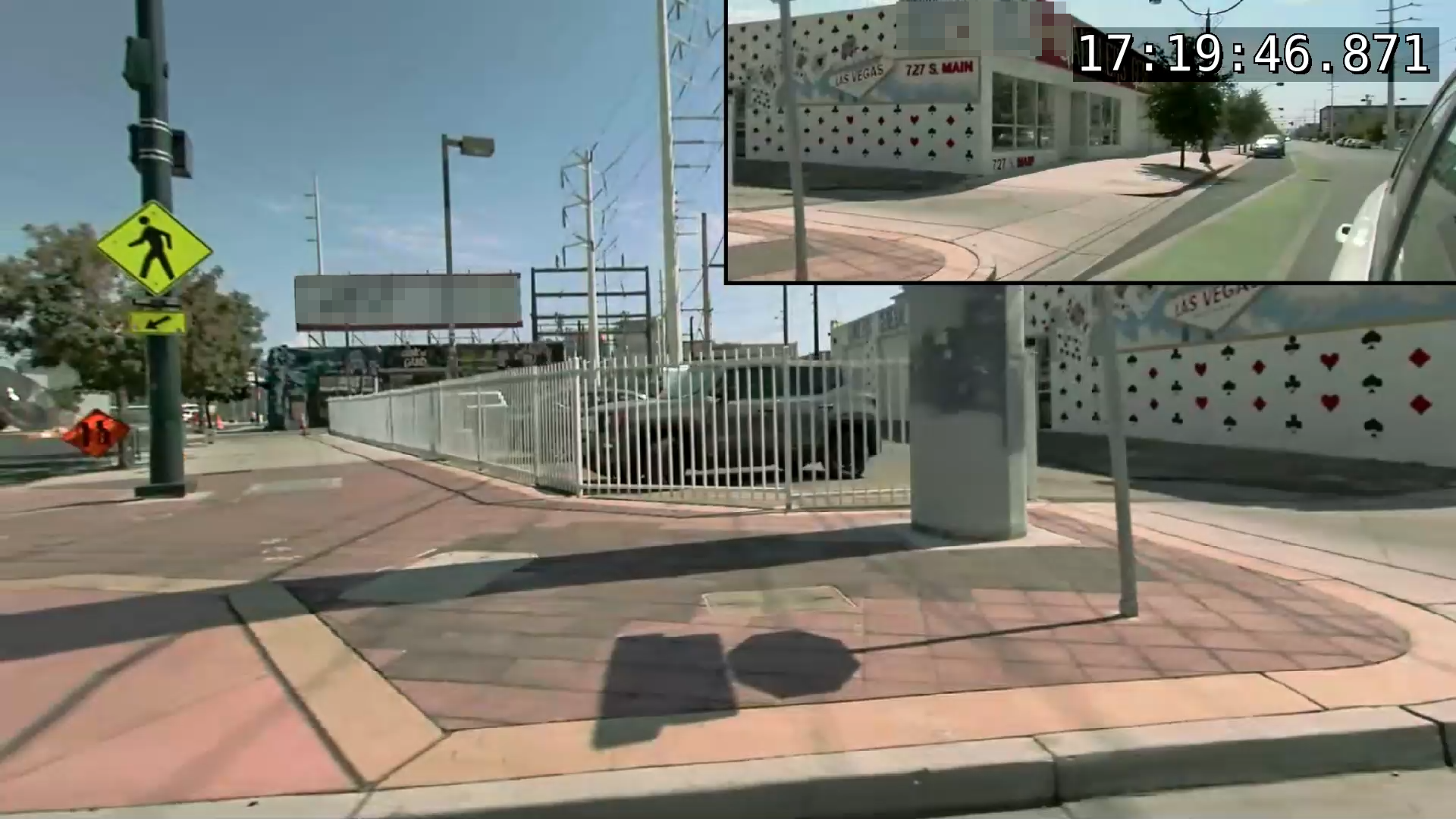}
        \caption{Right side screen}
        \label{Right_side_screen}
    \end{subfigure}
    \caption{Human Machine Interface (HMI) of the Remote Control Station (RCS) showing a right turn maneuver.}
    \label{Fig_HMI_Interface}
\end{figure*}

\section{REMOTE DRIVING SYSTEM MODEL}
\label{sec_systemmodel}
\begin{figure}[ht]
    \centerline{\includegraphics[width=2.0in]{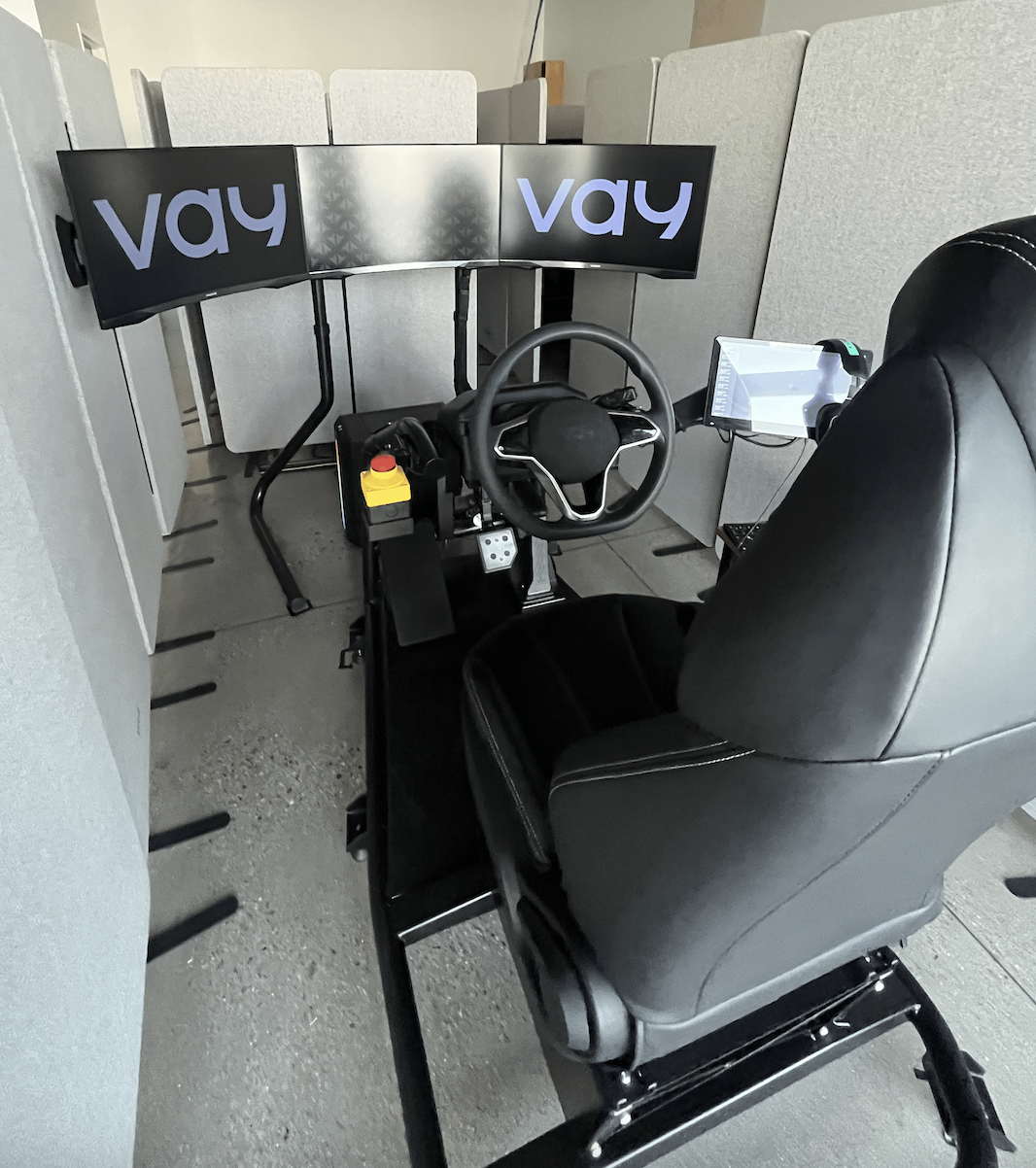}}
    \caption{Vay Remote Control Station (RCS) within an operations center in Las Vegas, Nevada, US.
    \label{Fig_VayRCS}}
    \vspace{-3mm}
\end{figure}

The RDS used for this work was developed by Vay Technology GmbH and operates in an ODD in Las Vegas, Nevada, US. The RDS consists of the vehicle, which is retrofitted with additional Vay hardware and software, the RCS, as visualized in Fig. \ref{Fig_VayRCS}, which allows for remote control, and the RD, the human operator. Vay Technology retrofitted an Electric Vehicle (EV) Kia eNiro to integrate remote driving technology, which includes additional cameras and a safety controller that monitors and regulates relevant safety parameters in real time. Furthermore, the vehicle's connectivity was ensured by using customized antennas and modems that were linked with the Vay connectivity software stack, resulting in a stable and fast communication link between the vehicle and the RCS. The RCS functions as the HMI for the RD, who operates the vehicle remotely. It includes three screens that show visual feedback from the vehicle's cameras and other sensors. Additionally, the headset microphone provide acoustic feedback and communication to the interior of the vehicle. Multiple camera sensors transmit visual experience, while external microphones deliver road traffic sounds to the RD's headphones. The vehicle is controlled with an automotive-grade physical steering wheel, as well as automotive-grade controls like column switches, throttle, and brake pedals. The RCS includes special controllers that process incoming data and allow interaction with the vehicle's Vay system.

\subsection{Remote Driving System Human-Machine-Interface}
\label{HMI}
The RCS HMI, shown in \mbox{Fig. \ref{Fig_HMI_Interface}}, displays all necessary information about vehicle control and the environment in an intuitive and straightforward format. It combines visual aids and real-time video data to provide more precise control. The instrument panel displays provide vehicle status information, including indicator lights such as the check engine light, and allow for the quick diagnosis of technical faults. In addition, the HMI displays current speed, gear selection, system latency, driving mode, and remaining range to support efficient and controlled driving. The blue trajectory lines show the planned vehicle path, while a so-called safety corridor on the left and right sides of the trajectory adds an extra meter of distance to the surroundings. This serves e.g. as orientation for parked vehicles with doors that can be opened. In addition, the HMI displays visual indicators for lateral and longitudinal acceleration and deceleration g-force based on the change of speed. 

\subsection{Remote Driving System Operational Design Domain}
\label{sec_RDS_ODD}
Within the larger Operational Domain (OD), the ODD specifies the system's particular operational scope, including the operating conditions that apply to the \mbox{system \cite{rohne2022implementing}}. A fallback for the Dynamic Driving \mbox{Task (DDT)} must be activated when the system requirements derived from the ODD can no longer be met, either because of system limitations or technical failures. This can involve either starting an active degradation function or switching to a Minimal Risk \mbox{Condition (MRC)} to reduce risks during the transition e.g. if the latency value exceeds a defined threshold \mbox{transition \cite{SAEI, no2021157}}. A Minimum Risk \mbox{Maneuver (MRM)} is a fundamental safety function to keep control within the ODD limitations and initiates an MRC, which represents a lower risk condition. The choice of the described ODD ensures that the RDS guarantees the required connectivity and controllability within a defined \mbox{environment \cite{hans2024backedautonomy}}. \mbox{Hans et al. \cite{hans2023operational}} defined a thorough ODD qualification method for remotely driven vehicles in urban settings \cite{hans2023operational}. The surroundings, traffic patterns, speed limits, weather, and time of day are all taken into consideration when determining the ODD of the RDS. The RDS utilized in this research operates in an urban environment in Las Vegas, Nevada, US. The urban environment offers a variety of traffic and infrastructure conditions that the system has to cope with. However, the ODD excludes specific parameters such as the following:

\begin{itemize}
    \item \textbf{Speed limitation:} Streets in the defined ODD are limited to a maximum of \mbox{35 mph}. 
    \item \textbf{Weather conditions:} Specific conditions such as snow, ice and rain are excluded from the ODD in this study.
    \item \textbf{Time of day:} The use of RDS is limited to driving during daylight hours in this study.
    \item \textbf{Sufficient and stable connectivity:} As \mbox{Hans et al. \cite{hans2023operational}} point out in their methodology for the ODD definition, connectivity is an essential prerequisite for the operation of the RDS within the described ODD and must be sufficiently guaranteed for the reliable exchange of information about the driving environment. 
\end{itemize}

\subsection{Remote Driver Training}
\label{sec_RDTraining}
To address the inherent limitations of the RDS, as described in \mbox{Section II.\ref{sec_remotedriving}}, the RD needs to meet certain requirements. Scientific research emphasizes that remote operation of the system and responding to vehicle-specific control requirements requires additional skills from experienced \mbox{drivers \cite{hans2024human, Schwindt-Drews2024}}. The RD training program considered in this study was developed by \mbox{Hans et al. \cite{Hans2023Academy}} and ensures technical and driving preparation of the RDs. The training is divided into the two main training areas of RDS-specific and ODD-specific RD training in addition to other objective metrics which are taken into account:

RDS-specific RD training focuses on the controllability of the RDS. RDs are familiarized with the specific functions and operating features of the system to develop an in-depth understanding of its technical capabilities and limitations. This includes controlling the RDS and adapting to the inherent system limitations, as explained in \mbox{Section II.\ref{sec_remotedriving}}. 

The ODD-specific training focuses on the practical application of the RDS within the defined ODD from \mbox{Section III.\ref{sec_RDS_ODD}}. Here, the RDs learn to take into account the special characteristics of remote driving in an urban environment, such as areas with potentially weak network connectivity. As \mbox{Hardin et al. \cite{Hardin2024well}} stated, local knowledge of roads is essential to focus on the actual driving task and anticipate potentially dangerous areas, especially those with unstable network connectivity. This local knowledge helps RD to cope with safety-relevant situations and contributes to a smoother integration into the traffic flow.

\section{SAFETY DRIVER INVESTIGATION RESULTS}
\label{sec_SD_interventions_results}
Disengagements during remote driving, in which a SD overrides the control inputs of the RD by pressing the accelerator or brake pedal or by applying torque to the steering wheel. Similar studies have investigated disengagements mainly in the context of ADSs where the in-vehicle SD intervenes when the ADS cannot perform the driving task \mbox{autonomously \cite{cummings2024identifying, kalra2016driving}}.

\begin{figure}
\centerline{\includegraphics[width=3.0in]{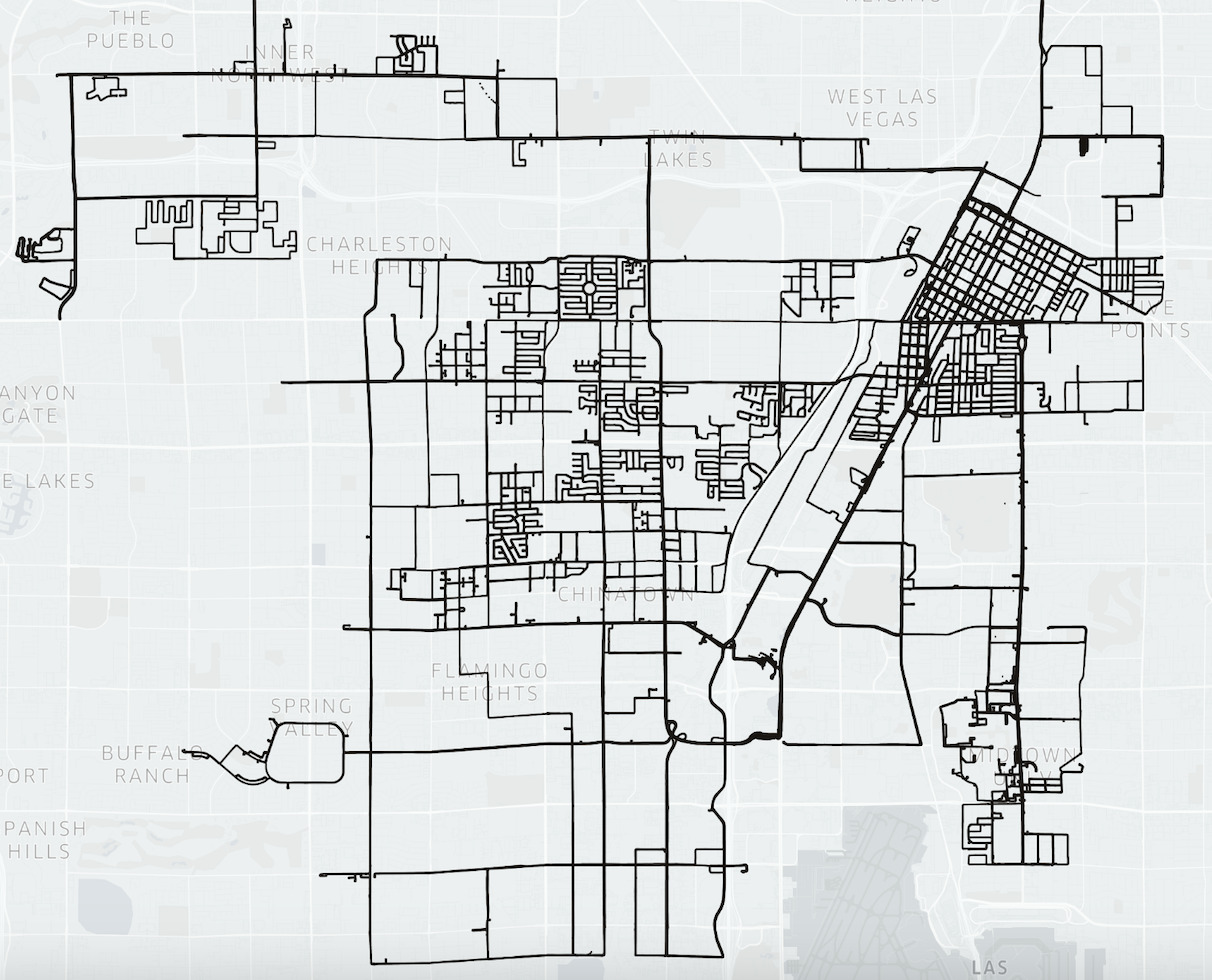}}
\caption{Extract of the geographical part of the Remote Driving System (RDS) Operational Design Domain (ODD) and distribution of the remotely driven distance of more than \mbox{14,000 km} in Las Vegas, Nevada, US.
\label{Fig_LV_ODD_extract}}
\vspace{-3mm}
\end{figure}

This Section of the study covers a total remote-controlled driving distance of \mbox{14,291.65 km}, spread over several driving sessions within the ODD defined in \mbox{Section III.\ref{sec_RDS_ODD}} and shown in \mbox{Fig. \ref{Fig_LV_ODD_extract}}, which was designed according to the ODD qualification process of \mbox{Hans et al. \cite{hans2023operational}}. As these session are RD training data sessions, an SD was present in all analyzed cases to ensure that control was taken if necessary.

In a period from \mbox{August 01, 2023}, to \mbox{December 01, 2024}, a total of 183 relevant disengagements were taken into account. Here, this work focuses exclusively on disengagements where the SD appropriately assumed control, such as in cases of errors made by the RD. The responsibility for these disengagements lies with the RD, whereby cases attributable to other traffic participant misbehavior or technical failures are excluded. Only disengagements where the RD did not react or did not react appropriately are analyzed. 

This part of the study includes a total of \mbox{25 RD}, aged between 22 and \mbox{44 years} (\mbox{$M = 30.44$}, \mbox{$SD = 4.81$}), where 21 identified as men and 4 identified as women. The data was filtered to include only designated RD in training, as defined in \mbox{Section III.\ref{sec_RDTraining}}, until a remote driving experience level of \mbox{800 km} and classified into the groups shown in \mbox{Tab. \ref{Tab_overview_td_level_disengagement}} with a comparable amount of km in the defined ODD. 

\begin{table}[ht]
\centering
\begin{tabular}{p{1.2cm}p{1.7cm}p{1.9cm}p{1.9cm}}
\hline
\textbf{RD Level} & \textbf{Driving \mbox{experience}} & \textbf{Remotely driven distance} & \textbf{Remotely driven duration}\\ \hline \hline
RD--L1 & \(<\)200 km    & 4132.56 km & 13,202.53 min\\ \hline
RD--L2 & 200--500 km  & 5494.06 km & 14,141.65 min\\ \hline
RD--L3 & 500--800 km  & 4665.04 km & 11,510.48 min\\ \hline \hline  
All Levels & 0--800 km & 14,291.65 km & 38,854.67 min\\ \hline 
\end{tabular}
\caption{Overview of remotely driven distance and remotely driven minutes by remote driving experience level.}
\label{Tab_overview_td_level_disengagement}
\vspace{-3mm}
\end{table}

\subsection{Number of SD Interventions over Driving Experience}
\label{sec_disengagementRDexperience}
\begin{figure}[ht]
\centerline{\includegraphics[width=3.3in]{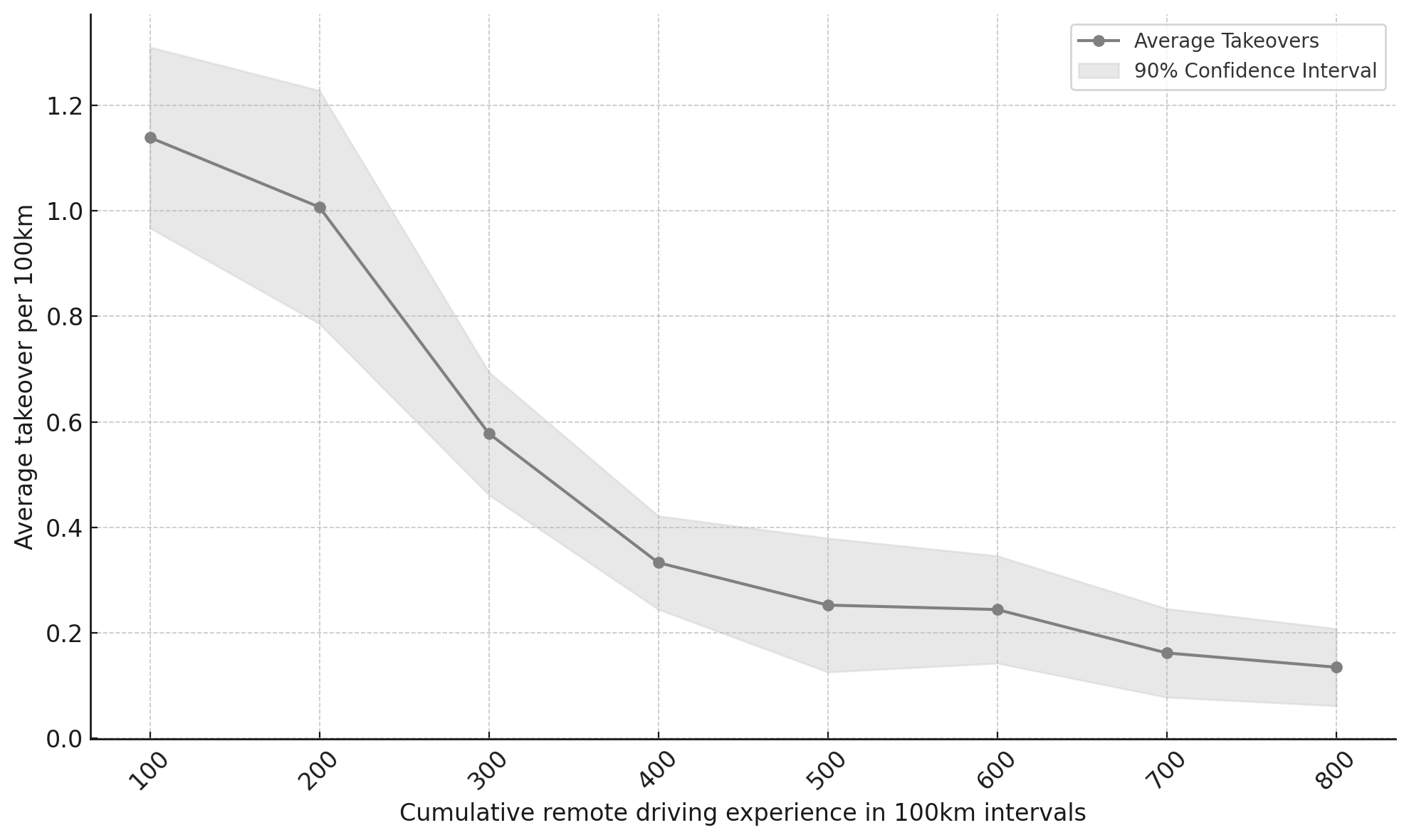}}
\caption{Evolution of the count of Safety Driver (SD) interventions over remote driving experience.
\label{Fig_disengagementRDexp}}
\end{figure}

Fig. \ref{Fig_disengagementRDexp} shows the correlation between the average number of SD interventions per \mbox{100 km} and the cumulative remote driving experience in intervals of \mbox{100 km}. The curve shows a clear decreasing trend of SD interventions per \mbox{100 km} with increasing cumulative remote driving experience. The data suggests that there is a pronounced learning curve for RDs. Especially in the first \mbox{400 km} of driving experience, a strong decrease in SD interventions can be seen, which could indicate a rapid acquisition of skills or an improved comprehension of the RDS. From about \mbox{400--500 km} the improvement in performance flattens out, so that additional experience only slightly reduces the number of interventions. This indicates that basic skills are developed at an early stage. The 90\% confidence interval is comparatively wide at low levels of experience, which indicates a greater dispersion of the data. This shows that individual performance varies greatly in terms of frequency of disengagements at low levels of remote driving experience. As experience increases, the confidence interval becomes narrower, indicating a more stable and consistent performance within the group of RDs. The increase in the confidence interval at around \mbox{500 km} can be attributed to a slightly higher number of SD interventions for individual RDs, which have a large effect on the confidence interval due to the small sample ($n=25$).

\subsection{Classification of SD Interventions}
\label{sec_disengagement_classification}
The various categories of SD interventions in relation to human RD performance are described below. The classification into these categories creates a basis for the quantitative and qualitative assessment of RD-related SD interventions. These are used for the detailed analysis of RD behavior and the identification of specific scenarios that can lead to potential safety-related situations.

\begin{itemize}
    \item \textbf{Braked late for signs:} The RD did not slow down the vehicle in time in front of a traffic sign.
    \item \textbf{Traffic light went red:} The RD did not stop the vehicle in time while the traffic light changed from green to red.
    \item \textbf{Impatient for 3rd parties, obstacles:} The RD did not react in time to respond to other traffic participants or objects.
    \item \textbf{Leaving the lane to the left:} The vehicle would have deviated from the intended lane to the left due to the RD's steering.
    \item \textbf{Leaving the lane to the right:} The vehicle would have deviated from the intended lane to the right due to the RD's steering.
    \item \textbf{Other:} This category includes other reasons for SD interventions due to the driving performance of the RD and do not fit into the above categories.
\end{itemize}

\begin{table*}[b]
\centering
\begin{tabular}{p{1.7cm} p{14.5cm}}
\hline
\textbf{Severity Level} & \textbf{Severity Classification Definition}\\ \hline \hline
SD-S3 & Significant chance of accident with injuries (life-threatening injuries and deaths).   \\ \hline
SD-S2 & Significant chance of accident with injuries (serious injuries, survival probable).    \\ \hline
SD-S1 & Significant chance of significant damage to the vehicle, but injuries unlikely or injuries are slight/moderate.   \\ \hline
SD-S0 & Significant chance of minor damage to vehicle (e.g. scratching, damage to mirror).  \\ \hline
SD-M2 & Car behaviour (would have) forced other traffic participants to react in a harsh/emergency maneuver, or a severe traffic violation would have resulted e.g. running a red light. \\ \hline
SD-M1 & Car behaviour (would have) forced other traffic participants to react, but no emergency maneuver from other traffic participants, or a non-severe traffic violation e.g. no stop in front of stop sign.\\ \hline
SD-M0 & Issue for the RD which affects the RD during driving, but from the outside, the car looks like it's behaving normally. \\ \hline
SD-U3 & Technically nothing went wrong, but passenger could have felt uncomfortable. \\ \hline
SD-U2 & Traffic situation requiring reaction outside of the usual flow of traffic. RD reacted well, situation could have been severe (SD-M2 or higher if RD had not reacted properly). \\ \hline
SD-U1 & Traffic situation requiring reaction outside of the usual flow of traffic. RD reacted well, situation was not severe (SD-M1 or less if RD had not reacted properly). \\ \hline
\end{tabular}
\caption{Severity classification for Safety Driver (SD) interventions specified for the remote driving application.}
\label{tab_severity_classification}
\end{table*}

\begin{figure}[ht]
\centerline{\includegraphics[width=3.3in]{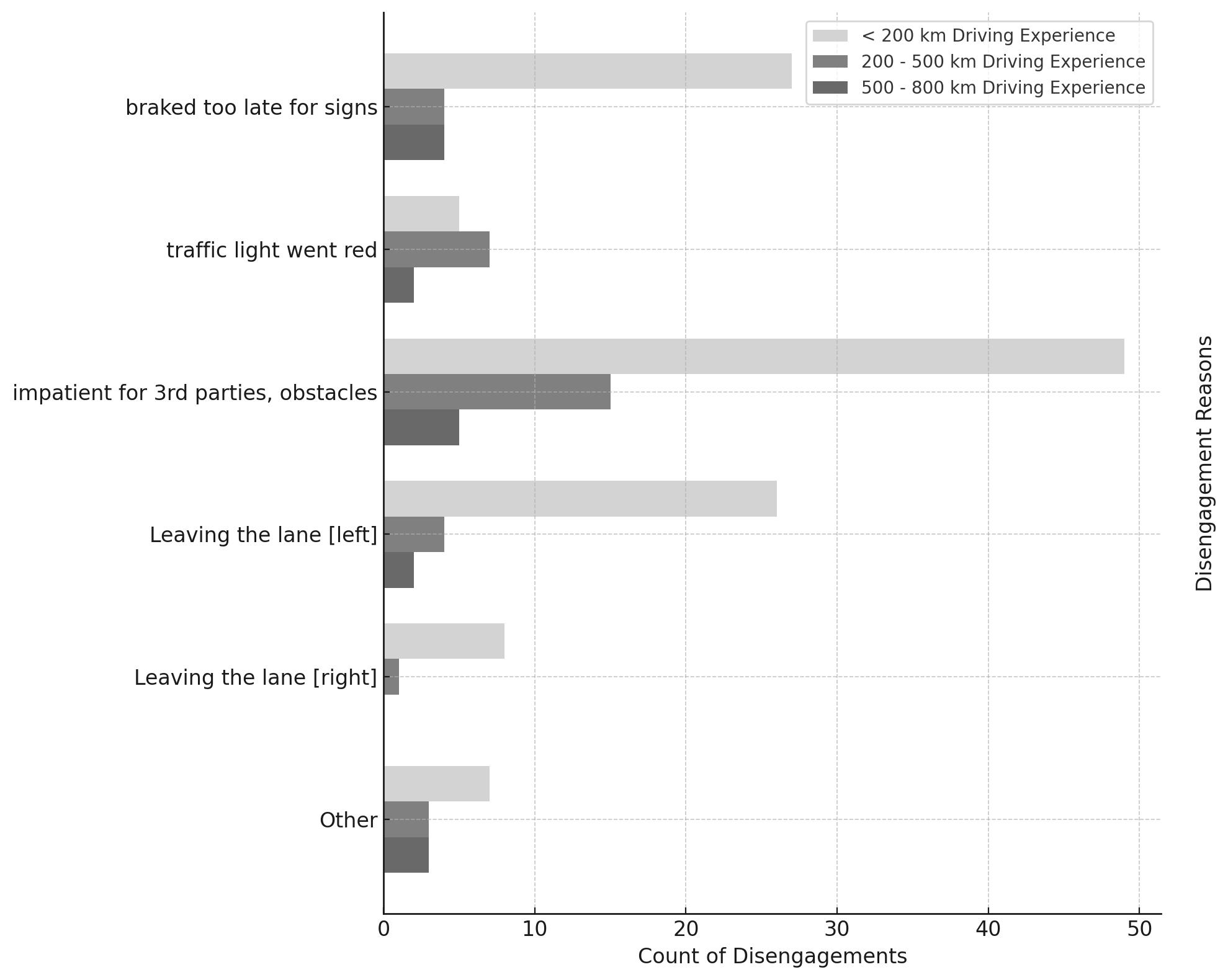}}
\caption{Count per disengagement reasons for different remote driving experience levels.
\label{Fig_disengagementreasons}}
\vspace{-5mm}
\end{figure}

\subsection{Distribution of Disengagement Reasons}
\label{sec_distribution_disengagement_reasons}
The distribution of disengagement into the reasons defined in \mbox{Section V.\ref{sec_disengagement_classification}} is shown in \mbox{Fig. \ref{Fig_disengagementreasons}}. The visualization supports the findings from \mbox{Section V.\ref{sec_disengagementRDexperience}} that there is a decreasing trend in the number of disengagements with increasing RD driving experience. Especially the difference between \mbox{RD--L1} and \mbox{RD--L2} illustrates that especially up to a driving experience of \mbox{200 km} there is the largest decrease in SD interventions, which is why within this range the RDs gain the most capabilities in controllability and improve the understanding of the RDS.

The results of the \mbox{Chi--Square} tests also support the findings of the Section V.\ref{sec_disengagementRDexperience}. \mbox{RD--L1} show significantly more frequent SD interventions in almost all categories, especially compared to \mbox{RD--L2} and \mbox{RD--L3}. This is clearly shown in the results of the Chi-Square test: A \(\chi^2\)--value of 21.70 (\(p = 0.006\)) was found between \mbox{RD--L1} and \mbox{RD--L2}, while the comparison between \mbox{RD--L1} and RD--L3 also showed significant differences (\(\chi^2 = 19.61\), \(p = 0.012\)). These results illustrate the significant learning progress associated with increasing remote driving experience. In contrast, the results of the \(\chi^2\)--test between \mbox{RD--L2} and \mbox{RD--L3} show no significant differences (\(\chi^2 = 5.70\), \(p = 0.681\)). This confirms that most of the improvement takes place in the first \mbox{500 km} of driving experience and that progress weakens thereafter. In combination with the significantly higher SD intervention rates of \mbox{RD--L1} in the categories \textit{impatient for 3rd parties, obstacles} and \textit{braked late for signs}, this illustrates that the critical phase of learning and adaptation to complex driving situations is in the initial phase of remote driving experience.

\subsection{SD Intervention Severity Classification for Remote Driving Applications}
\label{sec_severityclassificationresearch}
The classification of severity levels plays a central role in the assessment of risks in the context of road traffic in the automotive industry. The proposed severity classification from \mbox{Tab. \ref{tab_severity_classification}} is based on various established standards and legal regulations and adapted for remote driving purposes.

\begin{itemize}
    \item \textbf{SD-S0 to SD-S3:} The severity classes \mbox{SD-S0} to \mbox{SD-S3} were introduced on the basis of the severity classification of ISO 26262-3:2018 for functional safety of road vehicles \cite{ISO26262-3:2018}. ISO 26262 also takes into account the Abbreviated Injury Scale (AIS) \cite{haasper2010abbreviated}, a globally recognized classification system that rates injuries according to their severity. The decision to adopt this classification is based on the fact that both the standard and traffic accident statistics use a similar classification. This harmonization facilitates the comparability of analysis and safety assessments. 
    \item \textbf{SD-M1 and SD-M2:} The severity classes \mbox{SD-M1} and \mbox{SD-M2} are based on the basic principles of the German Road Traffic Regulations (StVO) \cite{StVO2013}. The aim of the StVO is to ensure a safe and smooth flow of traffic on public roads. In particular, the introduction of \mbox{SD-M1} and \mbox{SD-M2} was motivated by the two basic rules of §1 StVO.
        \begin{itemize}
        \item \textbf{SD-M1: Constant caution and mutual consideration:} Road users are obliged to behave prudently in order to minimize risks.
        \item \textbf{SD-M2: Avoidance of damage, danger or unavoidable obstruction:} Behavior in road traffic must be designed in such a way that no unnecessary burdens or dangers arise for others.
        \end{itemize}
    \item \textbf{SD-M0, SD-U1 to SD-U3:} When the SD adjusts the steering to maintain lane, avoid obstacles or react to sudden changes in the driving environment although the RD would have reacted correctly based on the telemetry data.
\end{itemize}

\subsection{Disengagements classified by Severity}
\label{disengagement_severity_application}
\begin{figure}[h!]
    \centerline{\includegraphics[width=3.25in]{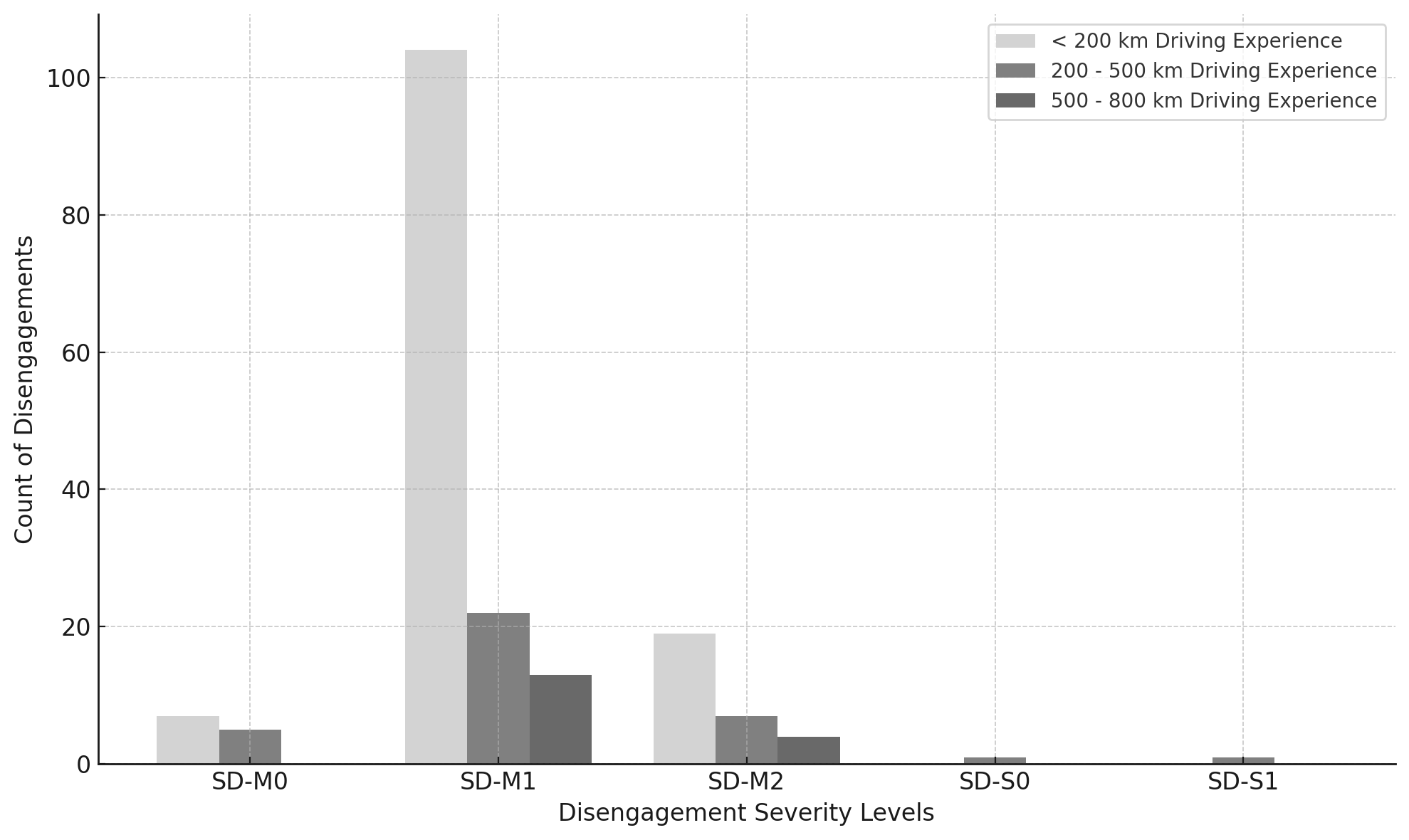}}
    \caption{Disengagement severity levels in relation to driving experience.
    \label{Fig_disengagementseverity}}
    \vspace{-3mm}
\end{figure}

Fig. \ref{Fig_disengagementseverity} shows the distribution of the number of disengagements depending on the severity category and the accumulated remote driving experience. The high number of disengagements in the \mbox{SD--M1} category \mbox{(104 cases)} is particularly noticeable, especially for RDs with less than \mbox{200 km} experience. With increasing driving experience, the number of SD interventions decreases significantly, which illustrates the learning curve. In the \mbox{SD--M2} category, disengagements also occur more frequently among inexperienced RDs \mbox{(22 cases)}, but are less frequent overall than in \mbox{SD--M1}. The most serious disengagements (\mbox{SD--S0} and \mbox{SD--S1}) are comparatively rare (1 case per category), which indicates an overall low number of potentially critical incidents that had to be prevented by the SD. 

The data suggests that as remote driving experience increases, the number of disengagements decreases overall, particularly in the \mbox{SD--M1} and \mbox{SD--M2} categories. \mbox{RD--L2} and \mbox{RD--L3} have significantly fewer disengagements than those with less than \mbox{200 km} experience. This confirms that remote control ability improves with increasing experience, as RD are increasingly able to adapt to inherent system limitations.

\section{RESULTS FOR HARSH AND EXTREME DRIVING EVENTS}
\label{sec_harsh_driving_events}
For a safe market introduction of remote driving technology, it is essential to prevent the necessity for SD interventions, ensuring that neither human nor technical disengagement requires manual compensation by an in-vehicle SD. To achieve this goal, it is necessary to identify potentially safety-related driving situations at an early stage. In particular, driving metrics must be used to recognize an increased probability of disengagements by the RD.

As part of this work, the RDS collects continuously real-time driving dynamics parameters, including vehicle position, speed and acceleration values. This information is transmitted through a telemetry system using proprietary driving physics messages to enable a detailed analysis of the vehicle dynamics. The analysis of these driving metrics focuses on key parameters such as longitudinal and lateral acceleration as well as braking behavior. These parameters form the basis for identifying specific driving events. The events are aggregated if they last for several seconds. In particular, events that occur within a time window of five seconds are aggregated to calculate the resulting event duration. This aggregation approach improves the accuracy of the driving style metrics and enables a differentiated analysis of longer driving scenarios. The study focuses on driving events with harsh and extreme parameter values defined in \mbox{Tab. \ref{motionparajournal}}, as these have an influence on driving behavior and vehicle dynamics. Harsh and extreme driving maneuvers are of particular interest here, as they represent potential indicators of increased disengagement risks. 

\begin{table}[h!]
\centering
\begin{tabular}{p{55pt}p{65pt}p{65pt}}
\hline
\textbf{Driving \mbox{Experience}} & \textbf{Remotely driven distance} & \textbf{Remotely driven duration}\\ \hline \hline
$<$ 500 km & 1311.26 km & 5004.23 min \\ \hline 
$>$ 500 km & 1333.22 km & 4417.82 min \\ \hline \hline
Total & 2644.48 km & 9422.05 min \\ \hline
\end{tabular}
\caption{Overview of Remote Driver (RD) experience for drivers with lower and higher remote driving experience than 500 km.}
\label{Tab_RDdisbalance}
\vspace{-3mm}
\end{table}

\begin{figure}[ht]
    \centerline{\includegraphics[width=1.8in]{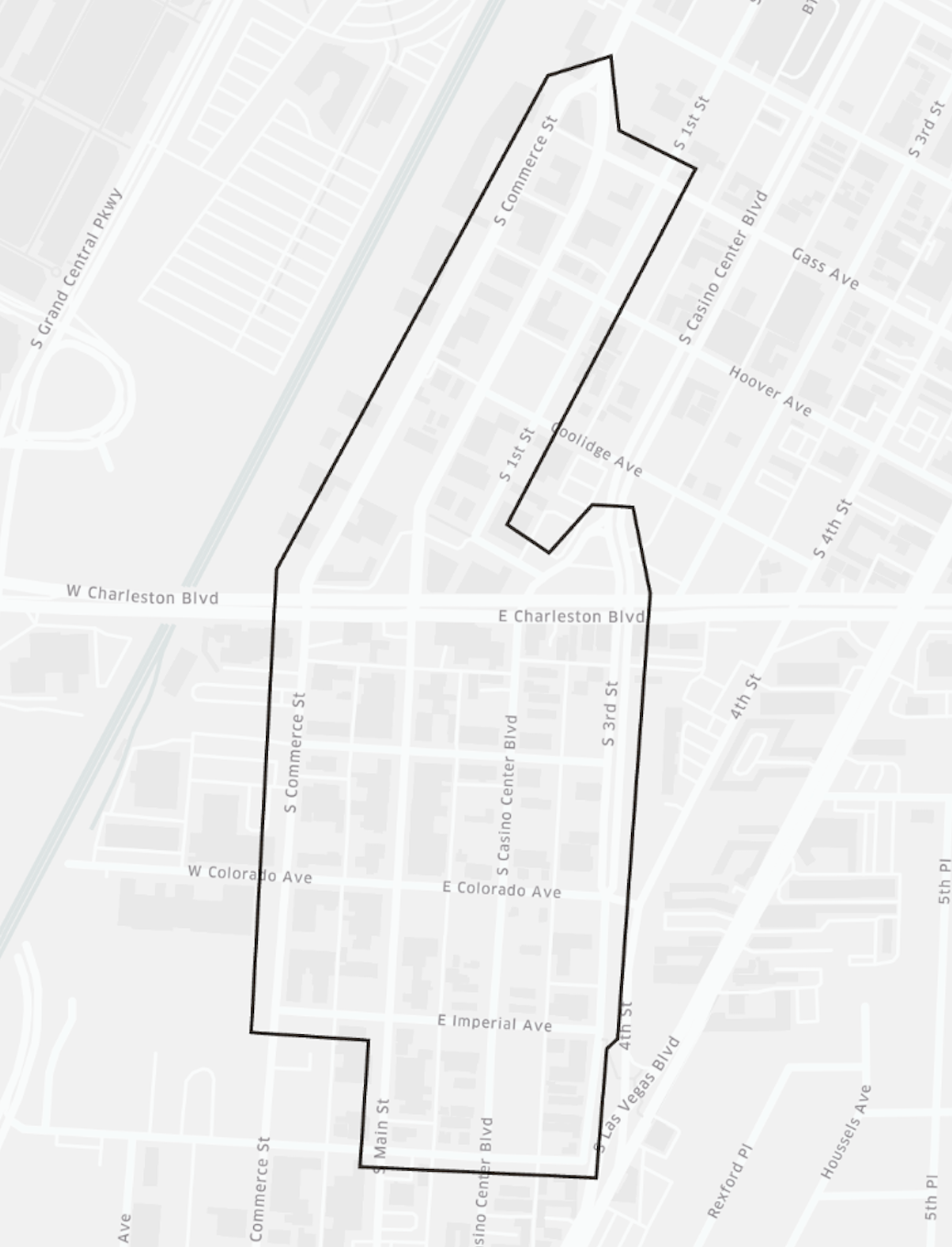}}
    \caption{Geographical part of the in Section \ref{sec_harsh_driving_events} considered Operational Design Domain (ODD) for harsh driving events in Las Vegas, Nevada, US. \label{Fig_ODD_Imperial_Ave}}
    \vspace{-1mm}
\end{figure}

The results and relevant data of this \mbox{Section \ref{sec_harsh_driving_events}} were collected over a period of one year from October, 30, 2023 to October, 30, 2024 in an ODD in Las Vegas, as shown in \mbox{Fig. \ref{Fig_ODD_Imperial_Ave}}. During this time period, a total of \mbox{2644.48 km} of remote driving data was collected in this ODD, which includes remote driving sessions with an SD as additional fallback during the training phase as well as remote driving sessions without a person in the vehicle. This driving data was collected from a total of 20 RDs with an average age between 22 and 44 years ($M = 31.65$, $SD = 5.49$), with 14 identifying themselves as men and 6 as women. These RD can be categorized into two experience groups, as seen in \mbox{Tab. \ref{Tab_RDdisbalance}}, which makes it possible to compare the results by remote driving experience.

\begin{figure}[ht]
\centerline{\includegraphics[width=3.3in]{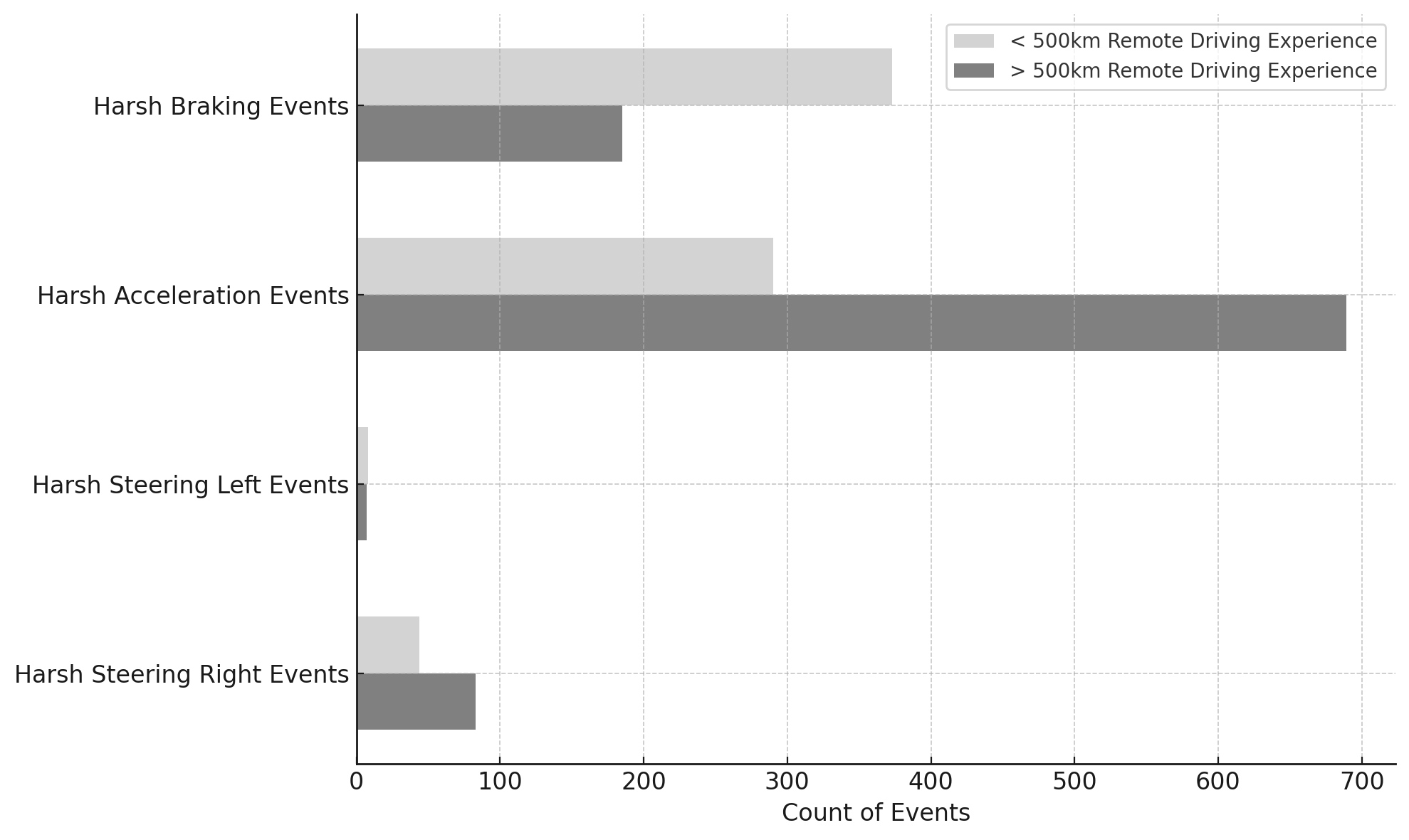}}
    \caption{Distribution and count of the relevant analyzed harsh driving events.
    \label{Fig_NumberEvents}}
    \vspace{-3mm}
\end{figure}

In total, 1646 relevant driving events were analyzed and classified, including 558 braking events, 979 acceleration events, 15 steering maneuvers to the left and 127 steering maneuvers to the right. The distribution of these events between the two experience groups defined in \mbox{Tab. \ref{Tab_RDdisbalance}} is illustrated in \mbox{Fig. \ref{Fig_NumberEvents}}.

\subsection{Harsh Braking Events}
\label{sec_harsh_braking_events}
A total of 558 relevant braking events were identified for further classification. These were classified into the five following categories, which enable a differentiated analysis of the harsh braking maneuvers:

\begin{itemize}
    \item \textbf{Impatient for other traffic participants:} The RD braked harsh to respond to other traffic participants or objects.
    \item \textbf{Traffic light went red:} The RD braked harsh to respond to the traffic light which changed from green to red.
    \item \textbf{Braked late for turn:} The RD braked harsh to ensure a smooth turning maneuver.
    \item \textbf{Braked late for signs:} The RD braked harsh to stop in time in front of a traffic sign.
    \item \textbf{Braked harsh for that event:} RD braked more than necessary for the specific scenario.
\end{itemize}

\begin{figure}[h!]
\centerline{\includegraphics[width=3.3in]{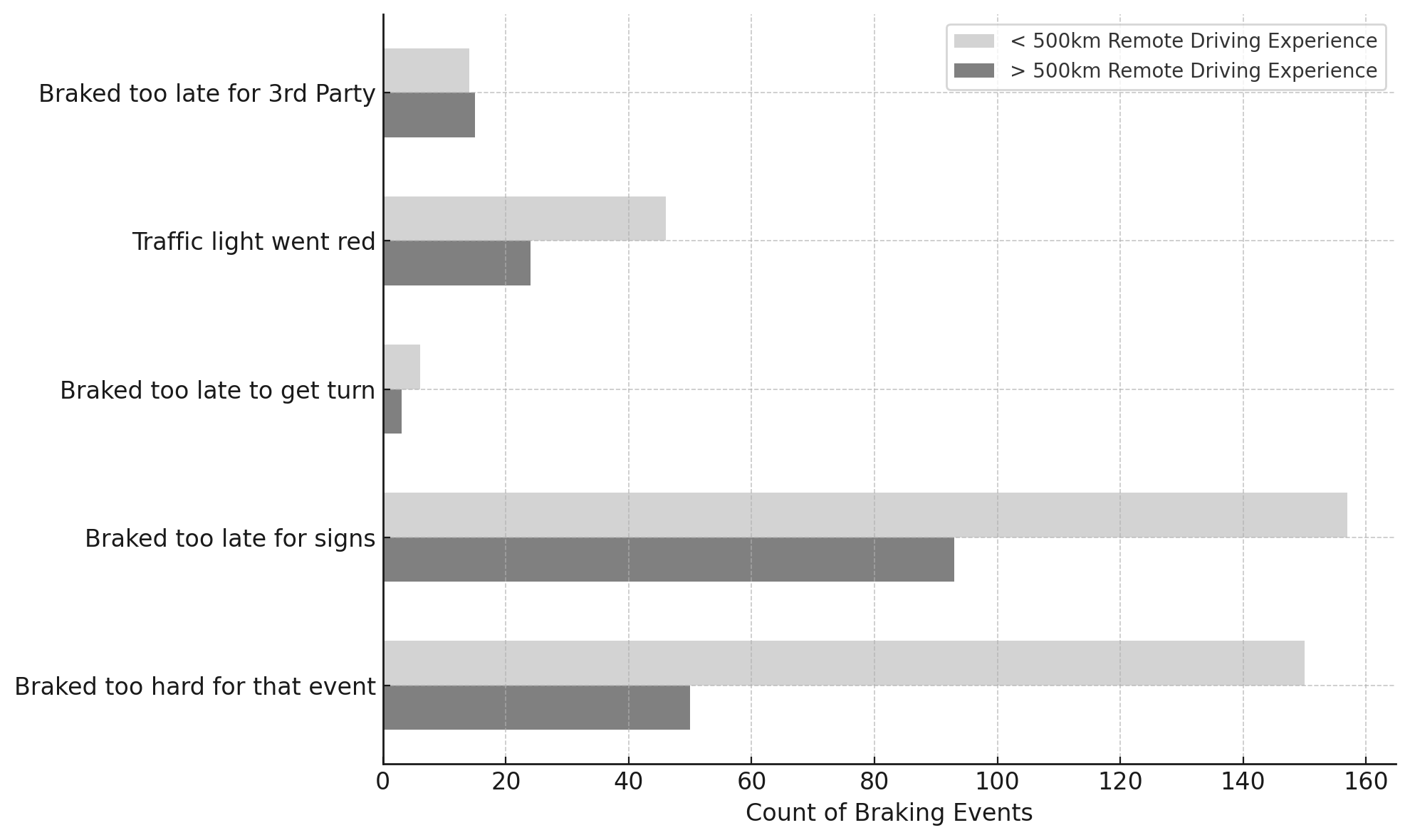}}
\caption{Distribution and count of harsh braking events per Remote Driver (RD) experience level.\label{Fig_DistributionBrakingEvents}}
\vspace{-3mm}
\end{figure}

\mbox{Fig. \ref{Fig_DistributionBrakingEvents}} shows the harsh braking events according to the classification of braking maneuvers with a subdivision according to driving experience into less than \mbox{500 km} RD experience and more than \mbox{500 km} RD experience. The differences between the two RD experience groups are statistically significant confirmed by the Chi-Square test \mbox{(\(\chi^2\) = 12.40, p = 0.015)}. The \mbox{\(\chi^2\)} value shows a moderate deviation between the observed and expected frequencies. Therefore, some categories contribute more to the statistically significant differences than others.

The analysis of the harsh braking scenarios shows two specific scenarios that demonstrate typical challenges in remote driving for the human RD. In the first scenario, the RD brakes late to stop in time in front of a stop sign or red traffic light, with the RD's initial speed exceeding \mbox{12 mph} in 88.8\% of cases, and even reaching a speed of \mbox{$\geq$ 19 mph} in 45.2\% of cases. In addition, this leads to a so-called \textit{sudden stop} in 93.9\% of cases, in which the RD continuously slowed down sharply the vehicle until it comes to a standstill. This observation is further supported by the category \textit{Braked harsh for that event}, in which 66\% of events occur before intersections or junctions with stop signs or traffic lights. In these cases, the RD also brakes hard, but overcompensates, so that the RD brakes the vehicle harder than necessary. Here too, the initial speed is \mbox{$\geq$ 12 mph} in 75\% of cases, which underscores the crucial role of speed in these maneuvers.

In the second scenario, the RD reacts to a yellow or red light phase of a traffic signal with a harsh braking event, again at a speed of at least \mbox{12 mph}. This scenario occurred in 91.43\% of cases at changing traffic signals, with 84.29\% of these events resulting in a sudden stop.

Both scenarios show that braking on point at high speed is one of the major challenges in remote driving for the RD. Although the statistics in \mbox{Figure \ref{Fig_DistributionBrakingEvents}} show that the probability of these scenarios occurring decreases with increasing experience, the high relevance of speed and reaction time when performing braking maneuvers in remote driving is nevertheless unambiguous.

\subsection{Harsh Steering Events}
\label{sec_harsh_steering_events}
The evaluation results from \mbox{Section VI.\ref{sec_harsh_braking_events}} that braking up to a certain point at high speed is a challenge in remote driving are confirmed by the analysis of harsh steering events. In 62.68\% of cases, the RD entered a curve at an initial speed of $\geq$ 12 mph, triggering a harsh steering event based on the defined thresholds from \mbox{Tab. \ref{motionparajournal}}. In 40.45\% of these cases (25.35\% of the total cases), a correction of the steering movement by the RD was also necessary. The discrepancy in the distribution of steering maneuvers can be attributed to the design of the ODD as shown in \mbox{Fig. \ref{Fig_ODD_Imperial_Ave}}. This results from the fact that left turns were mainly performed on roads that were previously driven at a speed of over \mbox{19 mph}. 

\begin{figure}[b]
    \centerline{\includegraphics[width=2.5in]{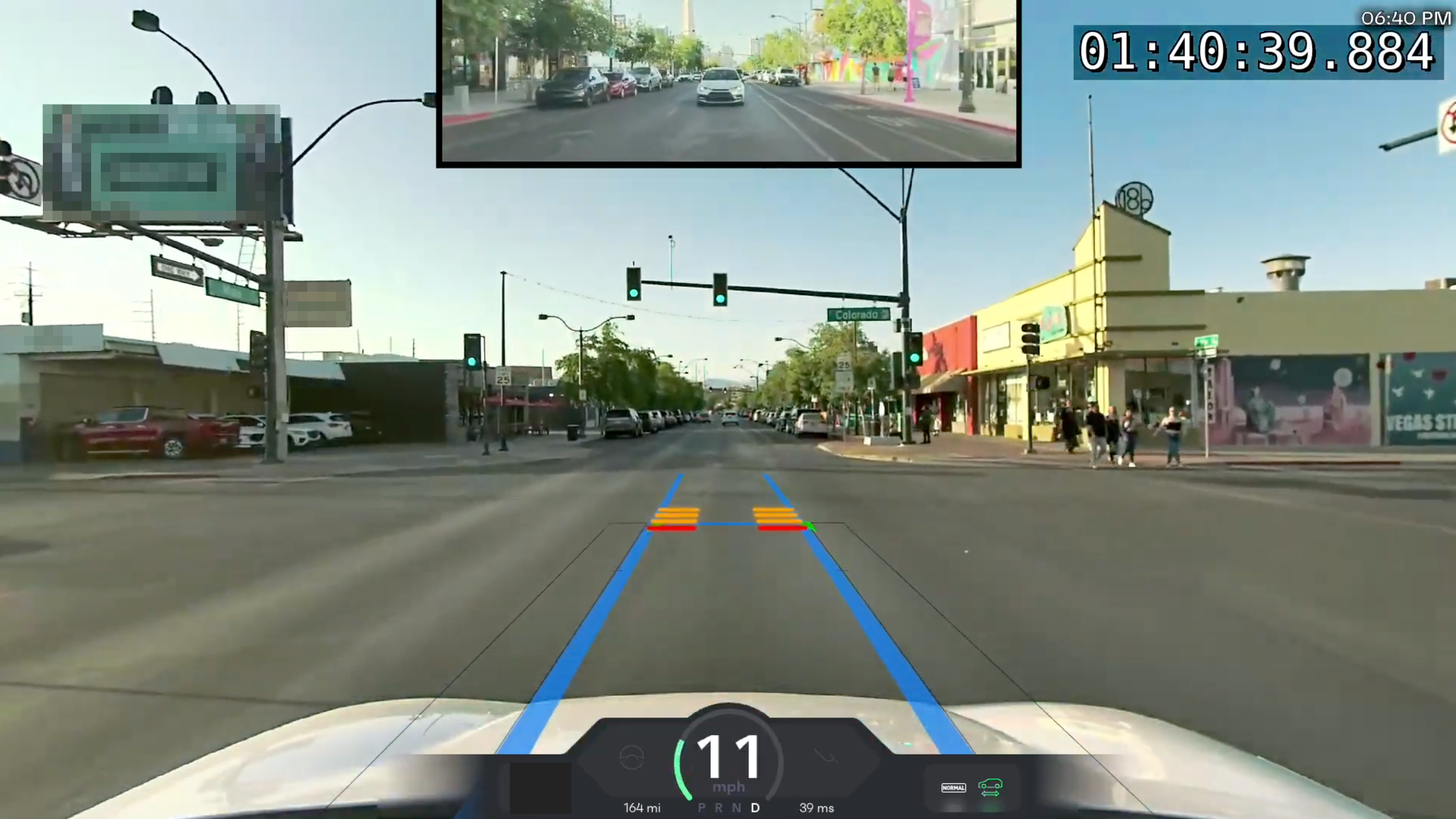}}
    \caption{Front screen Human Machine Interface (HMI) of the Remote Control Station (RCS) showing a harsh acceleration event.
    \label{Fig_harsh_acceleration}}
    \vspace{-3mm}
\end{figure}

\subsection{Harsh Acceleration Events}
\label{sec_harsh_acceleration_events}
The analysis of the 979 harsh acceleration events shows that in 66.72\% of these events, the RD accelerated harsh at a initial speed of less than \mbox{12 mph} before an intersection or junction. Of these events characterized by harsh acceleration, 65\% led to so-called \textit{sudden starts}. In these cases, the RD acceleration involves a continuous increase acceleration from standstill to at least \mbox{12 mph}. Such events indicate accelerated driving behavior, which may be due to an inadequately planned response by the RD. Sudden starts pose a particular challenge as they potentially affect the stability and safety level of the vehicle, especially in urban environments with variable traffic situations. In addition, 19.02\% of the harsh acceleration events were observed in the context of turning maneuvers. Within these maneuvers, 56.7\% of the events occurred in unprotected turns, accounting for 10.78\% of the total harsh acceleration events. Experienced RD have a higher number of maneuvers with high acceleration values, as they might have improved vehicle control and optimized decision-making. With increasing experience, uncertainty could be reduced, allowing dynamic maneuvers to be executed more precisely and efficiently. The two groups show only marginal differences in terms of median and mean acceleration, which indicates comparable central tendencies. However, \mbox{RD-L2} has a significantly lower standard deviation of 0.15 and a lower maximum value of $3.13 \text{m/s}^2$ than \mbox{RD-L1} (0.19 and $3.95 \text{m/s}^2$ respectively). This indicates a more consistent and less extreme acceleration behavior in \mbox{RD-L2}. The higher spread and the stronger outliers in \mbox{RD-L1} indicate a lower driving dynamic stability and potentially more impulsive inputs for less experienced drivers.

\section{RESULTS REMOTE DRIVER QUESTIONNAIRE}
\label{sec_results_questionaire}
The sample of RDs surveyed comprises 19 participants, mainly aged between 20 and 40 (94.7\%). Only one person is between 40 and 50 years old. The gender distribution shows a high proportion of 84.2\% of men (16), while 15.8\% of participants are female (3). In terms of conventional driving experience, 42.1\% of participants have less than 10 years experience (9), while 57.9\% have at least 10 years of driving experience (10). All participants have a car drivers license, 10.6\% have an additional motorcycle \mbox{license (2)} and 5.3\% have a truck drivers license (1). In terms of remote driving experience, 21\% of participants have driven up to 1000 km (4), which includes the training phase as well, 52.6\% between 1000 and \mbox{2000 km (10)}, and 26.3\% have more than 2000 km of \mbox{experience (5)}.

\subsection{Affinity for Technology Interaction (ATI)}
To measure technology affinity, the Affinity for Technology \mbox{Interaction (ATI)} Scale according to \mbox{Franke et al. \cite{franke2019personal}} was used. The mean value of the surveyed values is ($M = 4.51$) on a six-point scale, which indicates a high affinity for technology among the respondents. A value in the range of 4.1--5.0 indicates that the participants find technology-related interactions pleasant or helpful.

The standard deviation is $SD = 1.33$, which indicates a medium to high dispersion of values. Around 68\% of respondents have values in the range of 3.18 to 5.84 \mbox{(M ± SD)}. This shows that the majority of the sample is in the high technology affinity range, while some participants have a lower affinity. The calculated variance of $V = 1.78$ confirms a moderate heterogeneity in the responses. As this value is below 2.5, it can be ruled out that the sample is strongly polarized. The distribution of the values therefore shows no extreme differences within the group.

\subsection{Challenges with braking maneuvers during remote driving}
According to the participants, the biggest challenge in remote driving is braking maneuvers, which were described as challenging by 57.9\% of the participants. Difficulties arise in particular from the limited haptic feedback and the perception of distance and speed. 

One participant with less than \mbox{200 km} of remote driving experience emphasizes that "at high speeds and then braking down for a red traffic light is something a RD needs to get used to". Another participant with \mbox{200--500 km} of driving experience noted the difficulty in correctly estimating the deceleration of the vehicle, as "sometimes figuring out the braking pressure when coming to a stop can be difficult while trying not to harsh brake".

With increasing experience, RDs following defined rules helps to improve braking control. While drivers with less than \mbox{200 km} experience rely on early braking and speed reduction, more experienced drivers \mbox{(1000--1500 km)} use anticipatory driving or visual support on the screen such as visualization of stopping distance: "In order to stop correctly at stop lines I use my viz box on the stop line in tandem with intuitive braking". Another RD with the same experience level complemented: "I let go of the accelerator a few meters before I think I even have to brake". Very experienced RD with more than 3000 km remote driving experience report that braking becomes second nature over time: "[...] it's second nature now to know when to start feathering the brakes".

\subsection{Challenges when accelerating during remote driving}
Another challenging aspect is accelerating, reported by 15.8\% (2), in particular with regard to an appropriate dosage. RDs with 2000--2500 km experience report that the challenge is to accelerate from a standing start at the right speed rate - not too fast and not too slow. The high acceleration capacity of an EV in particular can cause unexpected reactions: "It can sometimes be easy to forget how quick EVs accelerate". A helpful tool here, as with braking, is the display of the longitudinal g-force on the screens of the RCS, which is intended to compensate for the limited haptic feedback with visual components.  

\subsection{Influence of experience on driving behavior}
The majority of 89.5\% of the RDs surveyed (17) stated that their remote driving experience and routine influence their driving behavior in the urban environment. Increasing familiarity with the system and the environment is emphasized particularly frequently: "Getting used to the system in detail and used to driving in a real world environment is definitely influenced by the driving experience". More experienced drivers with over 2000 km of driving experience see their remote driving practice as an important influencing factor for a controlled and driving style: "I drive safer/more defensive to others always being extremely aware of my surroundings". 

Only two participants (10.5\%) said that their experience had no noticeable impact on their driving behavior, with one person saying that it must become a second nature.

\subsection{Strategies to improve the perception of traffic participants}
Taking pedestrians and cyclists into account is a key challenge for RDs, as their movements are often unpredictable and require quick reactions especially in urban environment. With increasing experience in particular, RDs follow defined rules for proactively observing and anticipating potentially dangerous situations.  

RDs are increasingly focusing on proactive perception of traffic. A RD with \mbox{1000--1500 km} of remote driving experience emphasizes trying to “[...] anticipate movements of pedestrians and cyclists in advance”. An RD with \mbox{200--500 km} experience describes a similar approach and emphasizes the importance of an extended field of vision: “I look 15 seconds ahead and actively watch out for pedestrians and cyclists”. Very experienced RDs with a cumulative mileage of over \mbox{3000 km} emphasize the need for constant attention while driving. One RD in this experience group describes that RD "[...] always have to stay highly concentrated because you never know exactly what will happen next”. These statements make it clear that with increasing experience, not only anticipation but also continuous cognitive effort is seen as an essential component of controlled remote driving.

\section{LIMITATIONS}
\label{sec_limitation}
This study has several limitations that need be considered when interpreting the results:  

Firstly, this study has a limited and largely uniform sample size in the three analysis areas of this study. This means that outliers can have a stronger influence on the results. Nevertheless, this sample currently represents the largest available database of educated RD and allows the first in-depth insights into human performance within real-world ODDs. As a result, challenges in remote driving can be identified, confirmed or refuted for the first time.  

Secondly, the results of this work are based on a specific RDS used for passenger cars and a specific ODD. Therefore, different challenges may arise for other systems or vehicle types, such as trucks or vans, or in other ODD environments, such as in the desert or on highways. Nevertheless, the results provide a good indication of which problems are most likely to occur in these contexts. 

Finally, this study does not consider individual parameters of RDs. Factors such as situational awareness, cognitive perceptiveness or resilience to high workload can influence human performance in remote driving, as already mentioned in \mbox{Section II.\ref{sec_remotedriving}}. The small sample size also results in wider confidence intervals, as not every RD benefits equally from certain effects.

\section{CONCLUSION}
\label{sec_conclusion}
The results of this study show that remote driving is a demanding task that requires RD to adapt to the system and the traffic environment. The study shows that the first \mbox{200 km} of driving experience in particular play a crucial role in developing safe and efficient driving behavior. However, as experience increases, RDs develop increasingly effective strategies and are used to specific rules to overcome the various challenges, which is reflected in a significant reduction in SD interventions and a better response to critical traffic situations. The development of targeted training programs that address these challenges further increase the controllability and efficiency of remote driving systems.

The results of the SD interventions illustrate a clear learning curve. At the beginning of the remote driving experience, especially for RD with less than \mbox{200 km} experience, significantly more interventions occurred. These mainly concerned scenarios such as \textit{braking late in front of traffic signs} and \textit{impatient for other traffic participants}, which indicates a lack of adaptation to the system dynamics and insufficient vehicle control. The early phase makes it clear that the first \mbox{200 km} in particular represent a critical learning phase in which drivers acquire basic vehicle control skills and develop a better understanding of the system logic. With increasing experience, especially after the first 500 km, the number of SD interventions decreases significantly, indicating an improvement in control skills and a better understanding of the RDS. RDs with more experience reported improved control over the vehicle and a better understanding of the inherent challenges the system.

Harsh braking events proved to be one of the most frequent challenges in remote driving for the human RD. In particular, \textit{braking late before a stop sign} and \textit{braking hard in unexpected events} regularly led to interventions by the SD. This issue was particularly prevalent among RDs with less than \mbox{200 km} of experience, as braking force was applied late or appropriately. RDs with more experience reported following defined rules for anticipatory driving such as braking early and using visual aids and tools to assist in estimating braking distance. The application of these rules resulted in a lower frequency of SD interventions, reflecting the development of more effective vehicle control and increasing confidence in the system. Despite these improvements, the correct application of braking force, particularly at high speeds, remains a key challenge for the RD. This is underpinned by steering maneuvers, especially when cornering. In 62.68\% of cases where sharp steering maneuvers occurred, the speed at the start of the maneuver was over \mbox{12 mph}, requiring precise steering and often corrections by the RD. 

Harsh acceleration events were also a common challenge, especially the dosing of acceleration to control vehicle dynamics. For drivers with less than \mbox{200 km} of experience, improper acceleration metering led to harsh driving conditions, especially in urban environments with changing traffic situations. The high acceleration capacity of EVs used in the remote driving scenarios exacerbated this problem. RD with more experience reported that they were increasingly able to specifically control their acceleration behavior to overcome these challenges.

The results of the RD questionnaire underline the influence of RD experience on driving behavior. RD with more than 2000 km of experience reported that their driving style became more controlled as they developed a better understanding of the system and the environment. The importance of increasing familiarity with the system and the environment was particularly emphasized, resulting in a more proactive perception of road users and a more defensive driving style. Drivers with less experience, especially under \mbox{500 km}, pointed out that their reactions were often less precise, leading to an increased frequency of SD interventions.

The perception and anticipation of road users was another major challenge, especially with regard to pedestrians and cyclists, whose movements are often unpredictable. With increasing experience, RDs developed specific strategies, such as anticipating pedestrians' movements or widening their field of vision to anticipate potentially dangerous situations. These strategies, coupled with increased cognitive effort, were deemed necessary by the experienced drivers to ensure controllability during remote driving.

Future work should examine the development of RDs' skills over longer distances, especially after several thousand kilometers of driving experience. This should include analyzing how continuous experience affects the ability to anticipate and manage complex traffic situations. It would also be valuable to investigate whether there is a phase of plateauing in which additional experience brings about only minor improvements in driving behavior.

\section{ACKNOWLEDGMENTS}
This study was conducted supported by with Vay Technology, a company operating a remotely driven commercial car-sharing service. The analysis focuses on data collected from operations in Las Vegas, Nevada, US, while Vay Technology also conducts remote driving operations in Hamburg and Berlin, Germany. The authors would like to express sincere gratitude to Benedikt Walter and Hans-Leo Ross for their invaluable support and insightful input. 

In addition, the authors would like to thank Athanassios Lagospiris, Shilp Dixit, and Mathias Metzler for their constructive review and comments on this study. Their contributions were instrumental in shaping the direction and findings of this research.

\bibliography{Hans_main}

\begin{thebibliography}{10}
\providecommand{\url}[1]{#1}
\csname url@samestyle\endcsname
\providecommand{\newblock}{\relax}
\providecommand{\bibinfo}[2]{#2}
\providecommand{\BIBentrySTDinterwordspacing}{\spaceskip=0pt\relax}
\providecommand{\BIBentryALTinterwordstretchfactor}{4}
\providecommand{\BIBentryALTinterwordspacing}{\spaceskip=\fontdimen2\font plus
\BIBentryALTinterwordstretchfactor\fontdimen3\font minus \fontdimen4\font\relax}
\providecommand{\BIBforeignlanguage}[2]{{%
\expandafter\ifx\csname l@#1\endcsname\relax
\typeout{** WARNING: IEEEtran.bst: No hyphenation pattern has been}%
\typeout{** loaded for the language `#1'. Using the pattern for}%
\typeout{** the default language instead.}%
\else
\language=\csname l@#1\endcsname
\fi
#2}}
\providecommand{\BIBdecl}{\relax}
\BIBdecl

\bibitem{hans2024backedautonomy}
O.~Hans and B.~Walter, ``{ODD} design for automated and remote driving systems: A path to remotely backed autonomy,'' in \emph{2024 IEEE the 9th International Conference on Intelligent Transportation Engineering (ICITE)}.\hskip 1em plus 0.5em minus 0.4em\relax IEEE, 2024.

\bibitem{BSIFlex1887}
{British Standards Institution}, ``{BSI Flex 1887: Human Factors for Remote Operation of Vehicles – Guide v1.0},'' May 2023.

\bibitem{TeleoperationReport}
\BIBentryALTinterwordspacing
{Arbeitsgruppe Forschungsbedarf Teleoperation}, ``{Abschlussbericht der Arbeitsgruppe "Forschungsbedarf Teleoperation“},'' Sep. 2023. [Online]. Available: \url{https://www.bast.de/DE/Publikationen/Fachveroeffentlichungen/Fahrzeugtechnik/Downloads-Links/TO.html}
\BIBentrySTDinterwordspacing

\bibitem{Kelkar2025}
\BIBentryALTinterwordspacing
A.~Kelkar, K.~Heineke, M.~Kellner, and A.-S. Smith, ``Remote-driving services: The next disruption in mobility innovation?'' \emph{McKinsey \& Company}, January 2025. [Online]. Available: \url{https://www.mckinsey.com/features/mckinsey-center-for-future-mobility/our-insights/remote-driving-services-the-next-disruption-in-mobility-innovation}
\BIBentrySTDinterwordspacing

\bibitem{Heineke2024}
\BIBentryALTinterwordspacing
K.~Heineke, M.~Kellner, A.-S. Smith, and M.~Rebmann, ``Are consumers ready for remote driving?'' \emph{McKinsey \& Company}, September 2024. [Online]. Available: \url{https://www.mckinsey.com/features/mckinsey-center-for-future-mobility/mckinsey-on-urban-mobility/are-consumers-ready-for-remote-driving}
\BIBentrySTDinterwordspacing

\bibitem{roedl_fernlenkverordnung_2024}
{Rödl \& Partner}, ``{Entwurf der Straßenverkehrs-Fernlenkverordnung steht auf wackeligen Füßen},'' 2024.

\bibitem{neumeier2019teleoperation}
S.~Neumeier, P.~Wintersberger, A.-K. Frison, A.~Becher, C.~Facchi, and A.~Riener, ``Teleoperation: The holy grail to solve problems of automated driving? sure, but latency matters,'' in \emph{Proceedings of the 11th Int. Conf. on Automotive User Interfaces and Interactive Vehicular Applications}, 2019, pp. 186--197.

\bibitem{den2022design}
J.~den Ouden, V.~Ho, T.~van~der Smagt, G.~Kakes, S.~Rommel, I.~Passchier, J.~Juza, and I.~Tafur~Monroy, ``Design and evaluation of remote driving architecture on 4g and 5g mobile networks,'' \emph{Frontiers in Future Transportation}, vol.~2, p. 801567, 2022.

\bibitem{tener2022driving}
F.~Tener and J.~Lanir, ``Driving from a distance: challenges and guidelines for autonomous vehicle teleoperation interfaces,'' in \emph{Proceedings of the 2022 CHI conference on human factors in computing systems}, 2022, pp. 1--13.

\bibitem{UNECE}
H.~F. in~International Regulations~for Automated Driving Systems (HF-IRADS), ``Human factors challenges of remote support and control: A position paper from {HF-IRADS},'' vol.~8, pp. 1--9, 2020.

\bibitem{bogdoll2022taxonomy}
D.~Bogdoll, S.~Orf, L.~T{\"o}ttel, and J.~M. Z{\"o}llner, ``Taxonomy and survey on remote human input systems for driving automation systems,'' in \emph{Advances in Information and Communication: Proc. of the Future of Information and Communication Conference (FICC)}, 2022, pp. 94--108.

\bibitem{SAEI}
{On-Road Automated Driving committee}, ``Taxonomy and definitions for terms related to driving automation systems for on-road motor vehicles,'' 2021.

\bibitem{majstorovic2022survey}
D.~Majstorovi{\'c}, S.~Hoffmann, F.~Pfab, A.~Schimpe, M.-M. Wolf, and F.~Diermeyer, ``Survey on teleoperation concepts for automated vehicles,'' in \emph{2022 IEEE international conference on systems, man, and cybernetics (SMC)}.\hskip 1em plus 0.5em minus 0.4em\relax IEEE, 2022, pp. 1290--1296.

\bibitem{kettwich2021teleoperation}
C.~Kettwich, A.~Schrank, and M.~Oehl, ``Teleoperation of highly automated vehicles in public transport: User-centered design of a human-machine interface for remote-operation and its expert usability evaluation,'' \emph{Multimodal Technologies and Interaction}, vol.~5, no.~5, p.~26, 2021.

\bibitem{chen2007human}
J.~Y. Chen, E.~C. Haas, and M.~J. Barnes, ``Human performance issues and user interface design for teleoperated robots,'' \emph{IEEE Transactions on Systems, Man, and Cybernetics, Part C (Applications and Reviews)}, vol.~37, no.~6, pp. 1231--1245, 2007.

\bibitem{hortal2019rehabilitation}
E.~Hortal and E.~Hortal, ``Rehabilitation robot system,'' \emph{Brain-Machine Interfaces for Assistance and Rehabilitation of People with Reduced Mobility}, pp. 49--67, 2019.

\bibitem{hans2024human}
O.~Hans, K.~Radlak, and J.~Adamy, ``Human-factor focused application of stpa to remotely driven vehicles,'' in \emph{2024 IEEE World Forum on Public Safety Technology (WFPST)}.\hskip 1em plus 0.5em minus 0.4em\relax IEEE, 2024, pp. 120--125.

\bibitem{fong2001vehicle}
T.~Fong and C.~Thorpe, ``Vehicle teleoperation interfaces,'' \emph{Autonomous robots}, vol.~11, pp. 9--18, 2001.

\bibitem{georg2020sensor}
J.-M. Georg, J.~Feiler, S.~Hoffmann, and F.~Diermeyer, ``Sensor and actuator latency during teleoperation of automated vehicles,'' in \emph{2020 IEEE Intelligent Vehicles Symposium (IV)}.\hskip 1em plus 0.5em minus 0.4em\relax IEEE, 2020, pp. 760--766.

\bibitem{georg2019adaptable}
J.-M. Georg and F.~Diermeyer, ``An adaptable and immersive real time interface for resolving system limitations of automated vehicles with teleoperation,'' in \emph{2019 IEEE international conference on systems, man and cybernetics (SMC)}.\hskip 1em plus 0.5em minus 0.4em\relax IEEE, 2019, pp. 2659--2664.

\bibitem{hans2023operational}
O.~Hans, M.~Avezum, S.~Borysov, H.-L. Ross, and J.~Adamy, ``Operational design domain qualification framework for remotely driven vehicles in urban environment,'' in \emph{2023 IEEE International Automated Vehicle Validation Conference (IAVVC)}.\hskip 1em plus 0.5em minus 0.4em\relax IEEE, 2023, pp. 1--6.

\bibitem{Schwindt-Drews2024}
S.~Schwindt-Drews, J.~Frenzel, O.~Hans, B.~Abendroth, and J.~Adamy, ``Competencies and technological support for remote operator of automated vehicles: A literature review and perspectives,'' in \emph{AHFE 2024 International Conference Proceedings}, 2024.

\bibitem{martinez2017driving}
C.~M. Martinez, M.~Heucke, F.-Y. Wang, B.~Gao, and D.~Cao, ``Driving style recognition for intelligent vehicle control and advanced driver assistance: A survey,'' \emph{IEEE Transactions on Intelligent Transportation Systems}, vol.~19, no.~3, pp. 666--676, 2017.

\bibitem{elander1993behavioral}
J.~Elander, R.~West, and D.~French, ``Behavioral correlates of individual differences in road-traffic crash risk: An examination of methods and findings.'' \emph{Psychological bulletin}, vol. 113, no.~2, p. 279, 1993.

\bibitem{bellem2018comfort}
H.~Bellem, ``Comfort in automated driving: Analysis of driving style preference in automated driving,'' Ph.D. dissertation, Dissertation, Chemnitz, Technische Universit{\"a}t Chemnitz, 2018, 2018.

\bibitem{abendroth2009leistungsfahigkeit}
B.~Abendroth and R.~Bruder, \emph{Die Leistungsf{\"a}higkeit des Menschen f{\"u}r die Fahrzeugf{\"u}hrung}.\hskip 1em plus 0.5em minus 0.4em\relax Springer, 2009.

\bibitem{buyukyildiz2017identification}
G.~B{\"u}y{\"u}kyildiz, O.~Pion, C.~Hildebrandt, M.~Sedlmayr, R.~Henze, and F.~K{\"u}{\c{c}}{\"u}kay, ``Identification of the driving style for the adaptation of assistance systems,'' \emph{International Journal of Vehicle Autonomous Systems}, vol.~13, no.~3, pp. 244--260, 2017.

\bibitem{karjanto2017simulating}
J.~Karjanto, N.~Yusof, J.~Terken, F.~Delbressine, M.~Hassan, and G.~Rauterberg, ``Simulating autonomous driving styles: Accelerations for three road profiles,'' in \emph{MATEC web of conferences}, vol.~90.\hskip 1em plus 0.5em minus 0.4em\relax EDP Sciences, 2017, pp. 1--16.

\bibitem{schulz2008analyse}
A.~Schulz and R.~Fr{\"o}ming, ``{Analyse des Fahrerverhaltens zur Darstellung adaptiver Eingriffs-strategien von Assistenzsystemen},'' \emph{ATZ-Automobiltechnische Zeitschrift}, vol. 110, pp. 1124--1131, 2008.

\bibitem{pion2012fingerprint}
O.~Pion, R.~Henze, and F.~K{\"u}{\c{c}}{\"u}kay, ``{Fingerprint des Fahrers zur Adaption von Assistenzsystemen},'' in \emph{Informatik 2012}.\hskip 1em plus 0.5em minus 0.4em\relax Gesellschaft f{\"u}r Informatik eV, 2012, pp. 833--842.

\bibitem{corti2013quantitative}
A.~Corti, C.~Ongini, M.~Tanelli, and S.~M. Savaresi, ``Quantitative driving style estimation for energy-oriented applications in road vehicles,'' in \emph{2013 IEEE International Conference on Systems, Man, and Cybernetics}.\hskip 1em plus 0.5em minus 0.4em\relax IEEE, 2013, pp. 3710--3715.

\bibitem{manzoni2010driving}
V.~Manzoni, A.~Corti, P.~De~Luca, and S.~M. Savaresi, ``Driving style estimation via inertial measurements,'' in \emph{13th international IEEE conference on intelligent transportation systems}.\hskip 1em plus 0.5em minus 0.4em\relax IEEE, 2010, pp. 777--782.

\bibitem{guardiola2014modelling}
C.~Guardiola, B.~Pla, D.~Blanco-Rodr{\'\i}guez, and A.~Reig, ``Modelling driving behaviour and its impact on the energy management problem in hybrid electric vehicles,'' \emph{International Journal of Computer Mathematics}, vol.~91, no.~1, pp. 147--156, 2014.

\bibitem{schwab2019methode}
A.~Schwab, ``\BIBforeignlanguage{german}{{Eine Methode zur Auswahl kritischer Fahrszenarien für automatisierte Fahrzeuge anhand einer objektiven Charakterisierung des Fahrverhaltens}},'' mastersthesis, Technische Universität München, 2019.

\bibitem{lee2024study}
D.-W. Lee, T.-L. Kim, and S.-J. Kwon, ``A study on the driving performance analysis for autonomous vehicles through the real-road field operational test platform,'' \emph{International Journal of Precision Engineering and Manufacturing}, pp. 1--11, 2024.

\bibitem{mayser2004fahrerassistenzsysteme}
C.~Mayser, C.~Lippold, D.~Ebersbach, and M.~Dietze, ``{Fahrerassistenzsysteme zur Unterst{\"u}tzung der L{\"a}ngsregelung im ungebundenen Verkehr},'' in \emph{1. Tagung Aktive Sicherheit durch Fahrerassistenzsysteme}, 2004.

\bibitem{svensson2015tuning}
L.~Svensson and J.~Eriksson, ``Tuning for ride quality in autonomous vehicle: Application to linear quadratic path planning algorithm,'' 2015.

\bibitem{bosetti2014human}
P.~Bosetti, M.~Da~Lio, and A.~Saroldi, ``On the human control of vehicles: an experimental study of acceleration,'' \emph{European Transport Research Review}, vol.~6, pp. 157--170, 2014.

\bibitem{hugemann2003longitudinal}
W.~Hugemann and M.~Nickel, ``Longitudinal and lateral accelerations in normal day driving,'' in \emph{6th International Conference of The Institute of Traffic Accident Investigators}, 2003, pp. 1--8.

\bibitem{dorr2014online}
D.~D{\"o}rr, D.~Grabengiesser, and F.~Gauterin, ``Online driving style recognition using fuzzy logic,'' in \emph{17th international IEEE conference on intelligent transportation systems (ITSC)}.\hskip 1em plus 0.5em minus 0.4em\relax IEEE, 2014, pp. 1021--1026.

\bibitem{murphey2009driver}
Y.~L. Murphey, R.~Milton, and L.~Kiliaris, ``Driver's style classification using jerk analysis,'' in \emph{2009 IEEE workshop on computational intelligence in vehicles and vehicular systems}.\hskip 1em plus 0.5em minus 0.4em\relax IEEE, 2009, pp. 23--28.

\bibitem{kilinc2012determination}
A.~S. Kilinc and T.~Baybura, ``Determination of minimum horizontal curve radius used in the design of transportation structures, depending on the limit value of comfort criterion lateral jerk,'' \emph{TS06G-Engineering Surveying, Machine Control and Guidance}, 2012.

\bibitem{bae2019toward}
I.~Bae, J.~Moon, and J.~Seo, ``Toward a comfortable driving experience for a self-driving shuttle bus,'' \emph{Electronics}, vol.~8, no.~9, p. 943, 2019.

\bibitem{miyajima2007driver}
C.~Miyajima, Y.~Nishiwaki, K.~Ozawa, T.~Wakita, K.~Itou, K.~Takeda, and F.~Itakura, ``Driver modeling based on driving behavior and its evaluation in driver identification,'' \emph{Proceedings of the IEEE}, vol.~95, no.~2, pp. 427--437, 2007.

\bibitem{doshi2010examining}
A.~Doshi and M.~M. Trivedi, ``Examining the impact of driving style on the predictability and responsiveness of the driver: Real-world and simulator analysis,'' in \emph{2010 IEEE Intelligent Vehicles Symposium}.\hskip 1em plus 0.5em minus 0.4em\relax IEEE, 2010, pp. 232--237.

\bibitem{papaioannou2023motion}
G.~Papaioannou, L.~Zhao, M.~Nybacka, J.~Jerrelind, R.~Happee, and L.~Drugge, ``Motion comfort and driver feel: An explorative study about their relation in remote driving,'' \emph{arXiv preprint arXiv:2305.07370}, 2023.

\bibitem{rohne2022implementing}
D.~Rohne, A.~Richter, and E.~Schwalb, ``Implementing {ODD} as single point of knowledge to support the development of automated driving,'' in \emph{2022 IEEE Int. Conf. on Systems, Man, and Cybernetics (SMC)}.\hskip 1em plus 0.5em minus 0.4em\relax IEEE, 2022, pp. 1364--1370.

\bibitem{no2021157}
{{United Nations Economic Commission for Europe (UNECE)}}, ``Uniform provisions concerning the approval of vehicles with regard to automated lane keeping systems,'' \emph{Nations Economic Commission for Europe}, pp. 75--137, 2021.

\bibitem{Hans2023Academy}
O.~Hans, V.~Zaage, and I.~Zastrow, ``{Training for the backbone of our future mobility service: The Vay Teledrive Academy},'' Sep. 2023.

\bibitem{Hardin2024well}
B.~Hardin, P.~Salvini, M.~Jirotka, and L.~Kunze, ``How well do drivers adapt to remote operation? learning from remote drivers with on-road experience,'' 2024.

\bibitem{cummings2024identifying}
M.~Cummings and B.~Bauchwitz, ``Identifying research gaps through self-driving car data analysis,'' \emph{IEEE Transactions on Intelligent Vehicles}, 2024.

\bibitem{kalra2016driving}
N.~Kalra and S.~M. Paddock, ``Driving to safety: How many miles of driving would it take to demonstrate autonomous vehicle reliability?'' \emph{Transportation Research Part A: Policy and Practice}, vol.~94, pp. 182--193, 2016.

\bibitem{ISO26262-3:2018}
\emph{{ISO 26262-3:2018 - Road vehicles — Functional safety — Part 3: Concept phase}}, International Organization for Standardization Std., 2018.

\bibitem{haasper2010abbreviated}
C.~Haasper, M.~Junge, A.~Ernstberger, H.~Brehme, L.~Hannawald, C.~Langer, J.~Nehmzow, D.~Otte, U.~Sander, C.~Krettek \emph{et~al.}, ``Die abbreviated injury scale (ais): Potenzial und probleme bei der anwendung (leitthema),'' \emph{Der Unfallchirurg}, vol. 113, no.~5, pp. 366--372, 2010.

\bibitem{StVO2013}
\emph{Straßenverkehrs-Ordnung (StVO)}.\hskip 1em plus 0.5em minus 0.4em\relax Bundesministerium der Justiz und für Verbraucherschutz, 2013.

\bibitem{franke2019personal}
T.~Franke, C.~Attig, and D.~Wessel, ``A personal resource for technology interaction: development and validation of the affinity for technology interaction (ati) scale,'' \emph{International Journal of Human--Computer Interaction}, vol.~35, no.~6, pp. 456--467, 2019.

\end{thebibliography}

\begin{IEEEbiography}[{\includegraphics[width=1in,height=1.25in,clip,keepaspectratio]{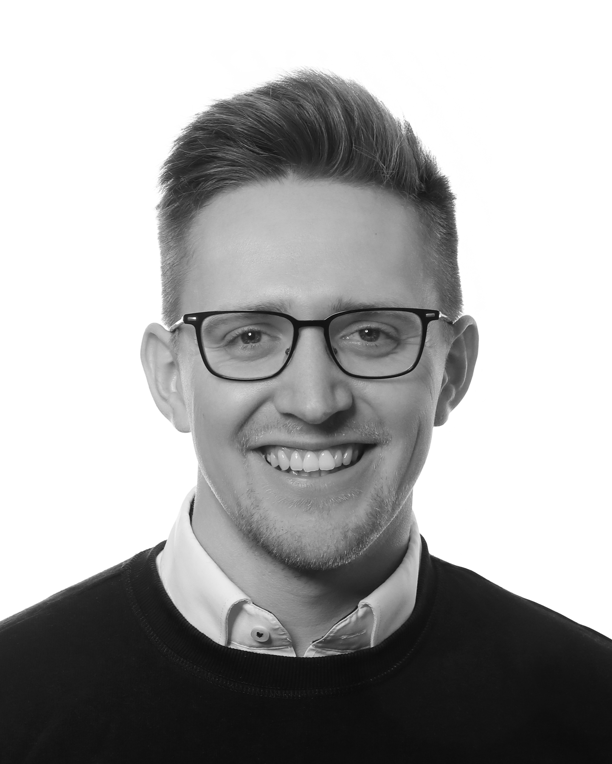}}]{OLE HANS } received the B.Sc. degree in Safety Engineering in 2020 and the M.Sc. degree in Quality Engineering in 2022 from University at Wuppertal. He is currently pursuing the Ph.D. (Dr.-Ing.) degree in electrical engineering and information technology with the Institute of Automatic Control and Mechatronics at the Technical University of Darmstadt. His research interests include safety of remotely driven and automated vehicles.
Since 2022, he has also been part of the Operational Safety Department at Vay Technology and is responsible for the safety-related operation of remote-driven vehicles.\end{IEEEbiography}

\begin{IEEEbiography}[{\includegraphics[width=1in,height=1.25in,clip,keepaspectratio]{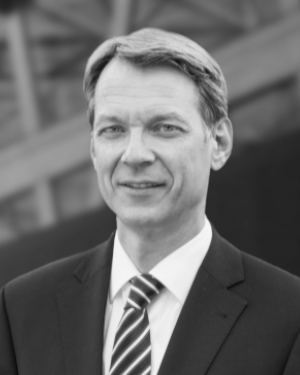}}]{JÜRGEN ADAMY }
received the Diploma and Dr.-Ing. degrees in Electrical Engineering from the Technical University of Dortmund, Germany, in 1987 and 1991, respectively. From 1992 to 1998, he was Engineer and Manager in the area of Control Applications at Siemens AG, Erlangen, Germany. In 1998, he became a Full Professor at the Technical University of Darmstadt, Germany, where he is the head of the Control Methods and Intelligent Systems Laboratory. His research interests are in the areas of nonlinear control, intelligent systems, and mobile robots.\end{IEEEbiography}

\end{document}